\documentclass[sigplan,10pt,screen]{acmart}
\usepackage{hyperref}

\usepackage[utf8x]{inputenc}
\usepackage{amsmath}
\usepackage{upgreek}
\usepackage{xcolor}
\usepackage{colortbl}
\usepackage[inline]{enumitem}
\usepackage[capitalise]{cleveref}
\usepackage{subcaption}

\usepackage{tikz-cd}
\usepackage{tabularx}

\definecolor{keywordcolor}{rgb}{0.7, 0.1, 0.1}   
\definecolor{tacticcolor}{rgb}{0.0, 0.1, 0.3}    
\definecolor{commentcolor}{rgb}{0.4, 0.4, 0.4}   
\definecolor{stringcolor}{rgb}{0.5, 0.3, 0.2}    
\definecolor{symbolcolor}{rgb}{0.1, 0.2, 0.7}    
\definecolor{sortcolor}{rgb}{0.1, 0.5, 0.1}      
\definecolor{attributecolor}{rgb}{0.7, 0.1, 0.1} 
\definecolor{errorcolor}{rgb}{1, 0, 0}           

\usepackage{listings}
\usepackage{flushend}
\usepackage{xspace}
\usepackage{underscore}
\usepackage{microtype}

\newcommand{\NN}{\mathbb{N}}
\newcommand{\ZZ}{\mathbb{Z}}
\newcommand{\QQ}{\mathbb{Q}}
\newcommand{\RR}{\mathbb{R}}

\newcommand{\Qp}{\mathbb{Q}_p}
\newcommand{\Zp}{\mathbb{Z}_p}
\newcommand{\mathlib}{\textsf{mathlib}\xspace}

\newcommand{\zmod}[1]{\ZZ/#1\ZZ}
\newcommand{\GF}[1]{\zmod{#1}}
\newcommand{\Witt}{\mathbb{W}}
\newcommand{\bind}{\textsf{bind}}
\renewcommand{\phi}{\varphi}

\newcommand*{\carry}[1][1]{\overset{#1}}
\newcolumntype{B}[1]{r*{#1}{@{}r}}

\newtheorem{stheorem}[equation]{Theorem}

\newtheorem{slemma}[equation]{Lemma}

\newcommand{\lean}[1]{\lstinline[language=lean]{#1}}

\usepackage{scalefnt}

\numberwithin{equation}{subsection}

\usepackage{etoolbox}
\newtoggle{COMMENTS}

\toggletrue{COMMENTS}

\setcopyright{none}
\copyrightyear{}
\acmYear{}
\acmDOI{}

\acmConference[Preprint]{Preprint}{2020}{}
\acmBooktitle{}
\acmPrice{}
\acmISBN{}

\begin{document}

\title{Formalizing the Ring of Witt Vectors}

\author{Johan Commelin}
\email{jmc@math.uni-freiburg.de}
\orcid{}

\affiliation{%
  \institution{Albert–Ludwigs-Universität Freiburg}
  \streetaddress{Ernst-Zermelo-Straße 1}
  \postcode{79104}
  \city{Freiburg}
  \country{Germany}
}

\author{Robert Y.\ Lewis}
\email{r.y.lewis@vu.nl}
\orcid{0000-0002-5266-1121}
\affiliation{%
  \institution{Vrije Universiteit Amsterdam}
  \streetaddress{De Boolelaan 1105}
  \postcode{1081 HV}
  \city{Amsterdam}
  \country{The Netherlands}}


\begin{abstract}
  The ring of Witt vectors $\mathbb{W} R$ over a base ring $R$ is an important tool
  in algebraic number theory
  and lies at the foundations of modern $p$-adic Hodge theory.
  $\mathbb{W} R$ has the interesting property that it constructs a ring of characteristic~$0$
  out of a ring of characteristic~$p > 1$,
  and it can be used more specifically to construct from a finite field
  containing $\mathbb{Z}/p\mathbb{Z}$
  the corresponding unramified field extension
  of the $p$-adic numbers~$\mathbb{Q}_p$
  (which is unique up to isomorphism).

  We formalize the notion of a Witt vector in the Lean proof assistant,
  along with the corresponding ring operations and other algebraic structure.
  We prove in Lean that, for prime $p$,
  the ring of Witt vectors over $\mathbb{Z}/p\mathbb{Z}$
  is isomorphic to the ring of $p$-adic integers $\mathbb{Z}_p$.
  In the process
  we develop idioms to cleanly handle calculations
  of identities between operations on the ring of Witt vectors.
  These calculations are intractable with a naive approach,
  and require a proof technique that is usually skimmed over in the informal literature.
  Our proofs resemble the informal arguments while being fully rigorous.
\end{abstract}

\ccsdesc[500]{Theory of computation~Type theory}
\ccsdesc[300]{Mathematics of computing~Mathematical analysis}
\ccsdesc[300]{Security and privacy~Logic and verification}

\keywords{formal math, ring theory, number theory, Lean, proof assistant}

\maketitle

\section{Introduction}

Formalizing a full undergraduate mathematics curriculum
has long been a goal of the proof assistant community~\cite{Wied07}.
This horizon is arguably now in sight:
most topics in the standard curriculum can be found
in at least one major proof assistant library.
As researchers, though, we cannot simply take this as a win.
With undergraduate mathematics done
we must turn to new challenges.

Formalizations of modern research mathematics are laudable,
but remain rare for good reason.
Such projects tend to take massive efforts~\cite{Gont13,Hales15},
to formalize only part of the main result~\cite{Stri19},
or to target theorems that are exceptionally well-suited for mechanization~\cite{Dahm19}.

This scarcity of results is hardly surprising.
Mastery of undergraduate topics is necessary to do research mathematics,
but far from sufficient:
Buzzard, Commelin, and Massot~\cite{BCM20} note the depth of theory that is needed
even to define the structures studied in many subfields.
We may be nearing the first horizon of undergraduate mathematics,
but the sea between us and the second horizon---graduate mathematics---is
vast, little explored, and filled with adventures.

As a new expedition into this sea,
we have constructed the ring of $p$-adic Witt vectors
and related operations
in the Lean proof assistant
and verified some of their fundamental properties.
Specifically, we define the Teichm\"uller lift
and the Frobenius and Verschiebung operators,
and show that the ring of Witt vectors over $\GF{p}$,
the integers modulo $p$,
is isomorphic to the $p$-adic integers.
To our knowledge, these topics have never before been formalized in a proof assistant.
Our development pushes forward the front line
of formalizations in ring theory.

Our project resulted in substantial additions to
the ring theory and multivariate polynomial sections
of Lean's mathematical library \mathlib~\cite{mathlib20}.
Building on Lewis' development of
the analytic properties of the $p$-adic numbers $\Qp$
and $p$-adic integers $\Zp$~\cite{Lewi19},
we have established more of their algebraic properties:
we show that $\Zp$ is a discrete valuation ring
and is the projective limit of the rings $\zmod{p^n}$ of integers modulo $p^n$.
Our project also served
to stress test Lean 3's type class inference mechanism in an algebraic context.

The early theory of Witt vectors was developed in the 1930s~\cite{Schm36,Witt37}.
They form a fundamental tool in algebraic number theory
and lie at the foundations of modern $p$-adic Hodge theory.
For example, they provide an elegant way to construct unramified $\Zp$-algebras
with prescribed finite residue fields of characteristic~$p$.
The ring of Witt vectors also appears
in the definitions of Fontaine's period ring $B_{\text{dR}}$~\cite{Fon94},
an important component in the classification of $p$-adic Galois representations.
Indeed, all the ingredients for the definition of $B_{\text{dR}}$
have now been formalized in Lean.

Witt vectors have a reputation among mathematicians
of being forbidding and impenetrable.
Presentations often skip the details of technical proofs and lengthy calculations;
these can become nightmarish unless approached very carefully.
One would reasonably expect a formalization to be even more nightmarish,
but we have found idioms in our development
that often lead to short, clean proofs and calculations,
clearer than their traditional counterparts.
In many cases, we have been able to reduce goals
to universal calculations in the language of rings~(\cref{universal-calculations-2}),
which can be discharged by very simple tactics~(\cref{subsection:auxtac}).
We believe that these statements and proofs are mathematically legible.
We were not able to erase the details in every case, though.
Our proofs that certain polynomials are integral
are long, slow, and unreadable,
just as they are on paper.

Our formalization is integrated into \mathlib.
We provide up-to-date information and links to the source code
on the project website:
\begin{center}\url{https://leanprover-community.github.io/witt-vectors}\end{center}

\section{Preliminaries}
\label{section:math-background}

The formalized contributions described in this paper can be roughly split into three parts:
\begin{enumerate}
  \item \label{contrib:alg} We expand the algebraic theory of the ring of $p$-adic integers $\Zp$
    (\cref{section:padics}).
  \item \label{contrib:wittdef} We define the notion of a Witt vector over an arbitrary ring $R$
    (\cref{section:polys})
    and construct a ring structure on the set of Witt vectors itself,
    additionally defining some fundamental operations on this ring
    (\cref{section:ring-operations}).
  \item \label{contrib:iso} We show that the ring of Witt vectors over $\GF{p}$ is isomorphic to $\Zp$
    (\cref{section:compare}).
\end{enumerate}
Parts~\ref{contrib:alg} and~\ref{contrib:wittdef} are independent of each other;
part~\ref{contrib:iso} bridges the first two.

To give the reader a high-level overview of the mathematical content of our formalization,
we sketch here the route that we will follow.
Since there is extensive introductory literature on the $p$-adic numbers
we focus on the latter parts.
Our main reference for part~\ref{contrib:alg} is Gouv\^ea~\cite{Gouv97},
although much is folklore.
Parts~\ref{contrib:wittdef} and~\ref{contrib:iso}
primarily follow Hazewinkel~\cite{Haze09}.

\subsection{$\Qp$ and $\Zp$}

The analytic perspective on the \emph{$p$-adic numbers} $\Qp$
defines them analogously to the real numbers $\RR$.
For a fixed prime number $p$,
$\Qp$ is the Cauchy completion of the rationals $\QQ$
with respect to the $p$-adic norm,
an alternative to the familiar absolute value
which is small for numbers whose numerators are divisible by large powers of $p$.
The field operations and norm on $\QQ$ lift to $\Qp$.
The \emph{$p$-adic integers} $\Zp$ are the $p$-adic numbers with norm at most 1;
they form a ring.

We can alternatively give an algebraic characterization of the $p$-adics.
From this perspective, we take $\Zp$ to be
the projective limit of $\zmod{p^n}$ in the category of rings
and $\Qp$ to be the field of fractions of $\Zp$.

Either perspective allows us to see $z \in \Zp$
as an infinite sum
$\sum_{k=0}^\infty z_k p^k$
where $z_k \in \ZZ$, $0 \leq z_k < p$ for each $k$.
(While this sum may diverge in the standard absolute value,
it always converges in the $p$-adic norm.)
This is particularly clear from the algebraic perspective,
as the $n$th partial sum corresponds to
an approximation to $z$ in $\zmod{p^n}$.
One can thus picture a $p$-adic integer
as a left-infinite base-$p$ expansion of digits (\cref{figure:arithmetic}).

\begin{figure}
  \begin{subfigure}[t!]{0.3\columnwidth}
  \begin{equation*}
 \begin{array}{B3}
     &\ldots  \mathtt{\carry 4\carry 4\carry 4\carry 4 \carry 4\carry 4\carry 4}&\mathtt{4} \\
       {} \hbox{$+$~} &                                                     &\mathtt{1} \\ \cline{2-3}
             &                                                    & \mathtt{0} \\
 \end{array}
 \end{equation*}
  \end{subfigure}
  ~
  \begin{subfigure}[t!]{0.3\columnwidth}
  \begin{equation*}
 \begin{array}{B3}
     &\ldots  \mathtt{\carry 3\carry[2] 1\carry 3 \carry[2] 1\carry 3\carry[2] 1 \carry 3}&\mathtt{2} \\
       {} \hbox{$\times$~}  &                                                   &\mathtt{3} \\ \cline{2-3}
             &                                                    & \mathtt{1} \\
 \end{array}
 \end{equation*}
  \end{subfigure}
 ~
  \begin{subfigure}[t!]{0.3\columnwidth}
  \begin{equation*}
 \begin{array}{B3}
     &\ldots  \mathtt{\carry 3\carry 1\carry 3 \carry 1\carry 3\carry 1 \carry 3}&\mathtt{2} \\
       {} \hbox{$+$~}  & \ldots \mathtt{4444444}                                           &\mathtt{4} \\ \cline{2-3}
             & \ldots     \mathtt{3131313}                                              & \mathtt{1} \\
 \end{array}
 \end{equation*}
  \end{subfigure}

  \caption{
    If we represent $\Zp$ as left-infinite streams of digits,
    we can perform addition and multiplication in base $p$
    by carrying remainders to the left.
    $5$-adically, $\ldots 444444 + 1 = 0$ and $\ldots 313132 \times 3 = 1$.
  }
  \label{figure:arithmetic}
\end{figure}

The $p$-adic numbers are fundamental
to many areas of number theory.
Among many other applications,
they appear in the studies of
Diophantine equations~\cite{lech1953}
and rational points on algebraic varieties~\cite{mccallum2012},
and lie at the core of the Hasse principle in
Diophantine geometry~\cite{browning2018}.

\subsection{The Ring of $p$-Typical Witt Vectors}
\label{witt-vectors}
\label{witt-polynomial} \label{ring-structure}

Fix a prime number~$p$ and a commutative ring~$R$.
The underlying set of the \emph{ring of $p$-typical Witt vectors $\Witt R$}
is the set of functions $\NN \to R$.
(Note that the prime number $p$ is usually suppressed in the notation~$\Witt R$.)
One usually pictures a Witt vector $x$ as a left-infinite
sequence of \emph{coefficients}:
\[
 (\dots, x_i, \dots, x_2, x_1, x_0), \quad x_i \in R.
\]
A very illustrative example to keep in mind is $\Witt (\GF{p})$,
in which the coefficients $x_i$ are integers modulo~$p$.
Readers may recognize the similarity to $\Zp$,
and we will eventually show that they are isomorphic as rings,
although this isomorphism is not the map that preserves the sequence of coefficients.

We will now describe some properties of~$\Witt R$.

First, $\Witt R$ is a commutative ring of characteristic~$0$,
even if $R$ has characteristic $p > 1$.
For Witt vectors $x = (\dots, x_1, x_0)$ and $y = (\dots, y_1, y_0)$ in $\Witt R$,
the addition and multiplication are defined as follows:
\begin{equation}
 \label{S-and-P}
\begin{aligned}
 x + y &= (\dots, S_i(x,y), \dots, S_1(x,y), S_0(x,y)) \\
 x \cdot y &= (\dots, P_i(x,y), \dots, P_1(x,y), P_0(x,y)) \\
\end{aligned}
\end{equation}
where the $S_i, P_i \in \ZZ[\dots, X_1, X_0, \dots, Y_1, Y_0]$
are certain polynomials that we will specify in \cref{subsection:polynomials}.
Importantly, these operations are not
the familiar componentwise addition and multiplication of sequences.
The $n$th entry, e.g. $S_n(x, y)$,
will depend on the entries $(x_n, \dots, x_0)$ and $(y_n, \dots, y_0)$
instead of only $x_n$ and~$y_n$.
This is similar to ``carrying'' arithmetic:
an overflow at one index creates a ripple that can reach arbitrarily far to the left.
It takes some machinery to establish that
these operations satisfy the axioms of a ring.

Second, $\Witt$ is functorial:
every ring homomorphism $f \colon R \to S$
induces a ring homomorphism $\Witt f \colon \Witt R \to \Witt S$
obtained by applying $f$ to all coefficients of~$x$.
This procedure preserves identity morphisms and compositions.

Third, we introduce the rings of truncated Witt vectors.
For a given natural number $n$,
one may truncate Witt vectors to their first $n$ coefficients,
which is compatible with the ring structure.
We therefore obtain a ring structure on $\Witt_n R = R^n$
and ring homomorphisms
\[
 \Witt R \to \Witt_n R, \quad x \mapsto (x_{n-1}, \dots, x_1, x_0).
\]
It is clear from this description
that $\Witt R$ is the projective limit of the rings $\Witt_n R$.
We describe this in more detail in \cref{subsection:truncated}.

Finally, for the purpose of this introduction,
there are several standard operations on Witt vectors
which in fact are natural transformations:
that is, they behave in the expected way with respect to the functoriality of~$\Witt$.
\begin{itemize}
 \item The Teichm\"uller lift is a multiplicative, zero-preserving map
  \[
   \tau \colon R \to \Witt R, \quad r \mapsto (\dots, 0, 0, r).
  \]
  In the example $\Witt (\GF{p}) \cong \Zp$,
  the elements $\tau(r) \in \Zp$ correspond to the $(p-1)$th
  roots of unity in $\Zp$ that can be obtained from~$\GF{p}$
  via Hensel's lemma (together with $\tau(0)=0$).

  \smallskip

 \item Verschiebung (``shift'') is an additive map
  \[
   V \colon \Witt R \to \Witt R, \quad x \mapsto (\dots, x_2, x_1, x_0, 0).
  \]

  \smallskip

 \item Frobenius is a ring homomorphism
  \[
   F \colon \Witt R \to \Witt R
  \]
  that is defined for general rings~$R$
  in a somewhat convoluted way.
  Suffice it to say that if $R$ is a ring of characteristic~$p$,
  then $f \colon R \to R, r \mapsto r^p$ is a ring homomorphism
  (also called Frobenius),
  and in this case $F = \Witt f$.

  \smallskip

 \item Multiplication by $n$ is denoted
  \[
   [n] \colon \Witt R \to \Witt R, \quad x \mapsto n \cdot x,
  \]
  and is an additive map.
\end{itemize}

These operations satisfy various identities
that we discuss in \cref{subsection:operators}.

\subsection{Universal Calculations}
\label{universal-calculations}

In the preceding section we have claimed various identities
of a ring-theoretic nature,
for example that addition and multiplication on the Witt vectors
are commutative and associative,
that the Teichm\"uller lifts are multiplicative,
and that Verschiebung is additive.
Direct approaches to proving these identities are bound to be messy,
to the point that they are futile.

We will now explain two strategies to approach the proofs of these relations
while containing the mess.
From a highbrow perspective, these strategies amount to the same thing,
but they are very different from the point of view of implementation
(both by hand and in Lean).
We apply both strategies in our formalization.

Before explaining these strategies,
we lay some groundwork that both have in common.
For $n \in \NN$, the $n$th \emph{Witt polynomial} is
\[
  W_n = \sum_{i = 0}^n p^i \cdot X_i^{p^{n-i}}
  \in \ZZ[\dots, X_1, X_0].
\]
(The Witt polynomials play a role in defining $S_i$ and $P_i$ in \cref{S-and-P}.)
If $x = (\dots, x_1, x_0) \in \Witt R$ is a Witt vector,
then $W_n(x) \in R$ is called the $n$th \emph{ghost component} of~$x$.
By definition of the ring structure on $\Witt R$,
this gives a ring homomorphism
\[
 w_n \colon \Witt R \to R, \quad x \mapsto W_n(x).
\]
These ghost components assemble into a ring homomorphism
called the \emph{ghost map}
\[
 w \colon \Witt R \to R^\NN, \quad x \mapsto (w_0(x), w_1(x), \dots),
\]
where the ring structure on the codomain
is given by pointwise addition and multiplication.
The ghost map is not injective in general,
but if $p$ is invertible in~$R$, then it is an isomorphism.

\paragraph{Strategy 1}
\begin{enumerate}
 \item First prove the identity for rings $R$ in which $p$ is invertible.
  Use the fact that $\Witt R$ is isomorphic to $R^\NN$ via the ghost map.
 \item Then prove the identity for polynomial rings over the integers:
  $R = \ZZ[(X_i)_{i \in I}]$.
  Use that these rings inject into
  $\QQ[(X_i)_{i \in I}]$, and apply the preceding point.
 \item Finally, use the natural surjective ring homomorphism
  \[
   \ZZ[(X_r)_{r \in R}] \to R, \qquad X_r \mapsto r
  \]
  to deduce the identity for arbitrary rings~$R$.
\end{enumerate}

\paragraph{Strategy 2 (sketch)}
\begin{enumerate}
 \item Ignore the fact that the ghost map is not injective in general.
 \item Apply the ghost map to both sides of the identity,
  and prove that the resulting claim is true in~$R^\NN$.
\end{enumerate}
Hazewinkel~\cite[p.14, footnote 14]{Haze09} writes of this strategy:
\begin{quote} 
 There are pitfalls in calculating with ghost components as is done here.
 Such a calculation gives a valid proof of an identity or something else
 only if it is a universal calculation; that is,
 makes no use of any properties beyond those that follow from the axioms
 for a unital commutative ring only.
\end{quote}

While Strategy 1 makes less of a mess than a naive approach,
it still opens some boxes better left closed.
Strategy 2 is enticing,
but it takes careful planning to make it amenable to formalization.
We discuss how we have done this in \cref{universal-calculations-2},
sidestepping the pitfalls that Hazewinkel warns about.
This strategy is a powerful method
for formally checking identities between the functions mentioned in
\cref{witt-vectors}:
with a simple Lean tactic for performing specific rewrites,
typical proofs take only two or three lines of code.

\subsection{Witt Vectors over $\GF{p}$}
\label{witt-fp}

We mentioned in \cref{witt-vectors} that $\Witt (\GF{p})$ is isomorphic to~$\Zp$.
This isomorphism is constructed in the following manner.
The ring $\Witt (\GF{p})$
is the projective limit of the rings of truncated Witt vectors~$\Witt_n (\GF{p})$.
Similarly, $\Zp$ is the projective limit of the rings~$\zmod{p^n}$.
It therefore suffices to construct isomorphisms $\Witt_n (\GF{p}) \to \zmod{p^n}$
that commute with the natural homomorphisms
\[
 \Witt_n (\GF{p}) \to \Witt_m (\GF{p}) \quad\text{and}\quad \zmod{p^n} \to \zmod{p^m}
\]
for all $m \le n$.
Since any two morphisms out of $\zmod{k}$ are always equal,
this commutativity condition is vacuously satisfied,
and we are left with constructing the isomorphisms $\Witt_n (\GF{p}) \to \zmod{p^n}$.
Using the fact that $\GF{p}$ has characteristic~$p$,
one can show that
\[
 p^i = (\dots, 0, 1, \underbrace{0, \dots, 0}_{\text{$i$ times}}) \in \Witt (\GF{p}).
\]
This means that for all $i < n$ we find $p^i \ne 0$ in $\Witt_n (\GF{p})$.
Hence $\Witt_n (\GF{p})$ is a ring of characteristic~$p^n$
that has cardinality~$p^n$.
It is therefore isomorphic to~$\zmod{p^n}$.
This completes the proof that $\Witt (\GF{p})$ is isomorphic to~$\Zp$.

\subsection{Lean and \mathlib}
\label{section:lean-background}

Our formalization is based on Lean's community-driven mathematical library \mathlib~\cite{mathlib20},
and the work we describe has been integrated into the library.
We depend on numerous modules in \mathlib that have been enhanced by earlier projects.
In particular, Lewis' construction of $\Zp$~\cite{Lewi19}
and preliminaries from
Buzzard, Commelin, and Massot~\cite{BCM20}
on the theory of valuation rings
serve as a solid foundation for our work.

We rely heavily on the theory of multivariate polynomials,
to which many community members have contributed.
The type \lean{mv_polynomial σ R},
where \lean{R} is a commutative semiring,
represents polynomials with coefficients in \lean{R}
whose variables are indexed by the type \lstinline{σ}.

Lean's core library and \mathlib are designed around
using \emph{type classes}~\cite{Wadl89,Spit11} to manage mathematical structure.
Our development takes this path as well.
Structures in \mathlib usually follow a \emph{partially bundled} approach,
where, for example, \lean{group G} is a \lean{Type}-valued predicate on a type \lean{G}
asserting that \lean{G} has a group structure.
While the group operations and their properties are bundled in the structure definition,
the carrier type \lean{G} is not.

An exception to this rule is \mathlib's use of bundled morphisms~\cite[Section 4.1.2]{mathlib20}.
The partially bundled approach would suggest to define a type class \lean{is_ring_hom f}
asserting that \lean{f : R → S} satisfies the properties of a ring homomorphism.
(The ring structures on \lean{R} and \lean{S} are provided by type class arguments.)
In practice, the issues with compositionality introduced by this approach
are worse than the problems it solves.
Instead, \mathlib defines a structure \lean{ring_hom R S},
with notation \lean{R →+* S},
that bundles a function \lean{R → S} with proofs that it satisfies the ring homomorphism properties.
A coercion from \lean{R →+* S} to \lean{R → S}
projecting out this function
allows us to apply ring homomorphisms as if they were native functions.
At first glance this may seem to cut against the grain of the type theory,
since superficially we do not work with native function types.
In practice it works without issue
and behaves predictably in its interactions
with Lean's type class inference and simplifier.
We use the same approach for ring isomorphisms \lean{R ≃+* S}.

Some Lean code snippets in this paper have been slightly edited for the sake of formatting.
We fix parameters \lean{p : ℕ} and \lean{R : Type} throughout,
assuming \lean{p} is prime and \lean{R} is a commutative ring.

\section{Algebra of $\Zp$}
\label{section:padics}

We begin with the \mathlib development of the $p$-adic numbers described by Lewis~\cite{Lewi19}.
This development defines $\Qp$ as the Cauchy completion of $\QQ$
with respect to the $p$-adic norm
and $\Zp$ as the subring of elements with norm at most~$1$.
It establishes some basic algebraic facts about $\Zp$,
including that it is a local ring with maximal ideal spanned by ${p}$.
Our goal is to further develop the algebraic theory of $\Zp$,
culminating in a proof of its universal property (\cref{figure:limit}),
that it is the projective limit of the rings $\zmod{p^n}$.

\begin{figure}
  \begin{tikzcd}[column sep=huge]
   & R \ar[dashed]{d} \ar[swap,"f_{n+1}",bend right]{ddl} \ar["f_n",bend left]{ddr} \\
   & \Zp \ar{dl} \ar{dr} \\
   \zmod{p^{n+1}} \ar["\text{mod}"]{rr} & & \zmod{p^n}
  \end{tikzcd}
  \caption{$\Zp$ is the projective limit of $\zmod{p^n}$.
  Any family of compatible morphisms $f_n \colon R \to \zmod{p^n}$
  factors uniquely through $\Zp$.
  }
  \label{figure:limit}
\end{figure}

We follow \mathlib in using the notation \lstinline{ℤ_[p]} for the Lean type \lean{padic_int p}.

\subsection{Algebraic Instances}
\label{subsection:algebraic-instances}

We first establish that $\Zp$ is a discrete valuation ring (DVR).
We will need to know something about the structure of the ideals of~$\Zp$.

In the interest of developing a full API,
we prove a number of lemmas characterizing open unit balls.
These are mostly variants of the following:
\begin{lstlisting}
lemma norm_le_pow_iff_mem_span_pow
  (x : ℤ_[p]) (n : ℕ) :
  ∥x∥ ≤ p ^ (-n : ℤ) ↔
    x ∈ (ideal.span {p ^ n} : ideal ℤ_[p])
\end{lstlisting}
The notation \lean{(t : T)} instructs Lean to elaborate \lean{t} with expected type \lean{T},
inserting coercions if necessary.
Ideals in \mathlib are not necessarily finitely generated;
an ideal of $R$ is an $R$-submodule of $R$,
with membership in an ideal defined to be membership in the carrier set of the submodule.

In addition to lifting the $p$-adic norm from $\QQ$ to $\Qp$ and $\Zp$,
it is also useful to lift the $p$-adic valuation $\nu_p$.
This was done by Buzzard, Commelin, and Massot~\cite{BCM20}
but not integrated into \mathlib.
We integrate their work and provide variants in terms of this valuation, e.g.
\begin{lstlisting}
lemma mem_span_pow_iff_le_valuation
  {x : ℤ_[p]} (hx : x ≠ 0) (n : ℕ) :
  x ∈ (ideal.span {p ^ n} : ideal ℤ_[p]) ↔
    ↑n ≤ x.valuation
\end{lstlisting}
Proving these results is straightforward.
The \lean{norm_cast} tactic~\cite{Lewi20},
developed to simplify expressions containing type coercions,
proved useful to handle the many embeddings between $\NN$, $\ZZ$, $\Qp$, and $\Zp$.

These various characterizations of the ideals of $\Zp$
and the fact that $p$ is prime in $\Zp$
are sufficient to show that $\Zp$ is a DVR.
Unfortunately DVRs provide an excellent example
of a familiar pitfall of formalization.
Wikipedia provides 11 equivalent characterizations of a DVR,
each one convenient in certain contexts,
but in a proof assistant we must choose one as primary.
We found that the existing \mathlib definition was not well suited to our application,
and had to develop an alternate characterization
and prove it equivalent to the existing criterion.

\subsection{Universal Property}
\label{subsection:padics:universal}

One can think of an element of $\Zp$
as a left-infinite base-$p$ expansion of numerals.
With this in mind, it is possible to visualize a map from $\Zp$ to $\zmod{p^k}$ for $k \in \NN$:
take the $k$ rightmost digits of the expansion.
It is perhaps harder to see how to define this on the analytic representation of $\Zp$
or that this operation is a ring homomorphism.

We define this family of homomorphisms recursively,
first handling the $k = 1$ case and then using this in the general case.
The definitions are similar, so we factor out a common constructor:
to produce a ring homomorphism \lstinline{ℤ_[p] →+*} \lstinline{zmod k},
it suffices to give \lean{f : ℤ_[p] → ℕ} satisfying certain properties.
Here, \lean{zmod k} is the \mathlib representation of $\zmod{k}$,
the ring of integers modulo \lean{k}.
\begin{lstlisting}
def to_zmod_hom (k : ℕ) (f : ℤ_[p] → ℕ)
  (f_spec : ∀ x,
    x - f x ∈ (ideal.span {k} : ideal ℤ_[p]))
  (f_congr : ∀ (x : ℤ_[p]) (a b : ℕ),
      x - a ∈ (ideal.span {k} : ideal ℤ_[p]) →
        x - b ∈ (ideal.span {k} : ideal ℤ_[p]) →
          (a : zmod k) = b) :
  ℤ_[p] →+* zmod k
\end{lstlisting}

Suppose $r \in \QQ$ with $\|r\|_p \leq 1$.
There is a unique integer $0 \leq m(p, r) < p$ such that
$\|r - m(p, r)\|_p < 1$.
Using that $\QQ$ is densely embedded in $\Qp$,
we can transfer this property from $\QQ$ to $\Qp$,
and rephrase using results from \cref{subsection:algebraic-instances}
as follows:
\begin{lstlisting}
lemma exists_mem_range (x : ℤ_[p]) :
  ∃ n : ℕ, n < p ∧ (x - n ∈ maximal_ideal ℤ_[p])
\end{lstlisting}
The function \lean{zmod_repr : ℤ_[p] → ℕ} projects out this value \lean{n}.
By construction, it satisfies the \lean{f_spec} requirement of \lean{to_zmod_hom},
and after a little more work to establish \lean{f_congr}
we can define \lean{to_zmod : ℤ_[p] →+* zmod p}.

For the general case,
we must define a family of functions
\lean{appr : ℤ_[p] → ℕ → ℕ}
such that \lean{appr x n} satisfies \lean{f_spec x} and \lean{f_congr x}
for \lean{k = p^n}.
These are effectively the ``$n$ rightmost digits'' functions mentioned above,
approximating \lean{x} to \lean{n} places.

The key to defining \lean{appr x} is to note that, for $x \neq 0$,
there is a unique unit element $u \in \Zp$ such that $x = u\cdot p^{|\nu_p(x)|}$.
We call this element \lean{unit_coeff x}.
We then define \lean{appr x n} by recursion on \lean{n : ℕ}.
\begin{lstlisting}
def appr : ℤ_[p] → ℕ → ℕ
| x 0     := 0
| x (n+1) :=
let y := x - appr x n in
if hy : y = 0 then appr x n
else let u := unit_coeff hy,
         v := |y.valuation - n|,
         d := to_zmod (u * (p ^ v)) in
  appr x n + p ^ n * d.val
\end{lstlisting}
In the recursive case,
we take \lean{y} to be the error in the previous approximation,
and apply \lean{to_zmod} to a product of \lean{unit_coeff y}.
This is the $(n+1)$th rightmost digit of our expansion,
so we can scale it and add it to the previous approximation.
After proving the specification and congruence properties of \lean{appr},
we again use \lean{to_zmod_hom} to define:
\begin{lstlisting}
to_zmod_pow (n : ℕ) : ℤ_[p] →+* zmod (p ^ n)
\end{lstlisting}

The construction of \lean{appr} may sound like a complicated way
to define a function with an intuitively simple description,
and indeed it takes some work to establish \lean{f_spec} and \lean{f_congr}.
It would be drastically simplified
if we began with an algebraic definition of $\Zp$
instead of the analytic one.
However, the complexity might resurface
in other places.

These analytic results are on display in the final step of this section,
when we show that $\Zp$ is the projective limit of $\zmod{p^n}$.
For a fixed ring $R$,
we work with a family of ring homomorphisms
$f_k : R \to \zmod{p^k}$
which we assume to be \emph{compatible}:
for any $r$ and $k_1 \leq k_2$, $f_{k_1}(r) \equiv f_{k_2}(r) \mod p^{k_1}$.
For any $r$, the sequence $n \mapsto f_n(r) \in \ZZ$ is Cauchy in the $p$-adic norm,
and thus converges in $\Zp$.
Calculations show that this map $R \to \Zp$ is a ring homomorphism,
so we define:
\begin{lstlisting}
def lift (f : Π (k : ℕ), R →+* zmod (p ^ k))
  (f_compat : ∀ k1 k2 (hk : k1 ≤ k2),
    (zmod.cast_hom (pow_dvd_pow p hk)).comp (f k2)
      = f k1) :
  R →+* ℤ_[p]
\end{lstlisting}
We finally show that \lean{lift} is the unique function
satisfying the commutative diagram in \cref{figure:limit},
establishing the universal property of $\Zp$
as the projective limit of $\zmod{p^n}$.

This result will be essential in \cref{section:compare}.
There, we will prove that $\Witt (\GF{p})$ satisfies the same universal property,
and conclude that the two rings are isomorphic.
In the meantime we face the substantial task of defining $\Witt$ and its ring structure.

\section{Witt Polynomials and Vectors}
\label{section:polys}

We can now continue to work toward the definition of $\Witt$.
While the bare definition is very easy to state,
we will need some machinery to define its ring structure,
so we develop that machinery first.

\subsection{Monadic Approach to Polynomials}

Key to simplifying statements in the realm of universal calculations
is the monadic \lean{bind} operation on the type of polynomials.
We often need to evaluate polynomials on other polynomials,
and defining it (together with a good collection of simplification lemmas)
made many calculations straight-forward.

Given \lean{f : σ → mv_polynomial τ R},
we define an algebra homomorphism
\begin{lstlisting}
bind₁ f :
  mv_polynomial σ R →ₐ[R] mv_polynomial τ R
\end{lstlisting}
that evaluates a polynomial in variables of type \lstinline{σ}
by sending each variable to its image under \lean{f}.
The subscript \lstinline{₁} distinguishes this from an analogous operation
that acts on the coefficient ring instead of the variables,
but we do not use \lean{bind₂} in our current development.

The \lean{bind₁} operator appears in many of our definitions and specifications,
and interacts naturally with the various Witt vector operations.
We register these interactions as simplification lemmas,
meaning that Lean's \lean{simp} tactic will by default use them to rewrite.
One can think of \lean{bind₁} as an atom in the universal language of rings:
when calculating, the definition \lean{bind₁} should never be unfolded,
and once other definitions are unfolded to the \lean{bind₁} level
the simplifier can often finish the calculation.

Note that in the informal notation, this operation is transparent,
and hence the calculations, involving say associativity of \lean{bind₁}
and renaming of variables, don't need to be performed either.
For our informal presentation here
we will denote the function \lean{bind₁ f} by $\bind_f$.

The \lean{bind} operator does indeed
induce a lawful monad structure on \lean{mv_polynomial}.
Its corresponding \lean{pure} operator
is the polynomial variable operator
\begin{lstlisting}
X : σ → mv_polynomial σ R
\end{lstlisting}
which lifts a term of the variable index type \lstinline{σ}
to a polynomial.
Its map operator, \lean{rename},
reindexes the variables via a map \lstinline{σ → τ}.

\subsection{Witt Polynomials and Structure Polynomials}
\label{subsection:polynomials}

We can now define the Witt polynomials,
which we will use to describe the ring structure on $\Witt R$.
Recall that for $n \in \NN$,
the $n$th \emph{Witt polynomial} is
\begin{equation}
  \label{equation:witt-poly}
 W_n = \sum_{i = 0}^n p^i \cdot X_i^{p^{n-i}}
 \in \ZZ[\dots, X_1, X_0].
\end{equation}
Their Lean definition is a direct translation of \cref{equation:witt-poly}:
\begin{lstlisting}[breaklines=false]
def witt_polynomial (n : ℕ) : mv_polynomial ℕ R :=
∑ i in range (n+1),
  monomial (single i (p ^ (n - i))) (p ^ i)
\end{lstlisting}
We use the notation \lean{W_ R n} for this type.

It is not so hard to see
that over the rationals, but not the integers,
the polynomials~$W_n$ form an alternative basis
of the polynomial algebra $\QQ[\dots, X_1, X_0]$,
so that by abuse of notation we may write
\[
 \QQ[\dots, W_1, W_0] \cong \QQ[\dots, X_1, X_0].
\]
In Lean, we define polynomials \lean{X_in_terms_of_W p R n}
that correspond to \lean{X n} viewed on the basis of Witt polynomials.
In other words,
applying \lean{bind₁ (W_ R)} to the polynomial \lean{X_in_terms_of_W p R n}
produces \lean{X n},
and similarly if we swap the polynomials.
This fact is key to establishing the algebra automorphism
that makes it easy to prove the following lemma.
For reasons of exposition,
we only treat the case where $\Phi$ is a polynomial in two variables,
but apart from notational complexity the case of an arbitrary (even infinite)
number of variables is not different at all.%
\begin{slemma}
 \label{witt-structure-rat-exists-unique}
 Let $\Phi \in \QQ[X,Y]$ be a polynomial.
 Then there exists a unique sequence of polynomials
 \[
  \phi_n \in \QQ[\dots, X_1, X_0, \dots Y_1, Y_0], \quad (n \in \NN)
 \]
 such that for all natural numbers $n$
 \[
  W_n(\dots \phi_1, \phi_0) = \Phi(W_n, W_n).
 \]
\end{slemma}
The monadic \lean{bind₁} makes another appearance
in the formal statement of this lemma:

\begin{lstlisting}
theorem witt_structure_rat_exists_unique
  (Φ : mv_polynomial idx ℚ) :
  ∃! (φ : ℕ → mv_polynomial (idx × ℕ) ℚ),
    ∀ (n : ℕ), bind₁ φ (W_ ℚ n) =
      bind₁ (λ i, (rename (prod.mk i) (W_ ℚ n))) Φ
\end{lstlisting}

A non-trivial calculation shows that if $\Phi$ has integral coefficients,
then so do the $\phi_n$.
Thus we get the following key theorem,
on which the whole theory of Witt vectors relies.
\begin{stheorem}
 \label{witt-structure-int-exists-unique}
 Let $\Phi \in \ZZ[X,Y]$ be a polynomial.
 Then there exists a unique sequence of polynomials
 \[
  \phi_n \in \ZZ[\dots, X_1, X_0, \dots Y_1, Y_0], \quad (n \in \NN)
 \]
 such that for all natural numbers $n$
 \[
  W_n(\dots \phi_1, \phi_0) = \Phi(W_n, W_n).
 \]
\end{stheorem}

The details of implementing this non-trivial calculation
are not pleasant,
involving arguments about the badly behaved numerator and denominator functions.
This is indeed one of the few points at which we step outside the language of rings.
The key ingredient in the proof is the following basic but non-trivial
number-theoretic fact.
\begin{lstlisting}
lemma dvd_sub_pow_of_dvd_sub {p : ℕ} {a b : R}
  (h : (p : R) ∣ a - b) (k : ℕ) :
  (p^(k+1) : R) ∣ a^(p^k) - b^(p^k)
\end{lstlisting}

Coq's Mathematical Components library~\cite{Mahb17}
provides an interface for manipulating polynomials
whose coefficients lie in a subring of a base ring.
There is no analogous interface for \mathlib's multivariate polynomials,
but in retrospect, it seems likely that this approach,
with base ring $\QQ$ and subring $\ZZ$, may have helped here.

The sequences of polynomials $S_n$ and~$P_n$
that occur in the definition (\cref{S-and-P})
of the addition and multiplication on~$\Witt R$
will be obtained by applying this theorem
to the polynomials $X + Y$ and $X \cdot Y$ respectively.
We explain in \cref{subsection:witt-lean}
why these operations satisfy the axioms of a commutative ring.

\subsection{The Type of $p$-Typical Witt Vectors}

After \cref{universal-calculations-2},
we will have all the necessary machinery to define a ring structure
and operations on $\Witt R$.
Before that, though, we must specify what a Witt vector actually is.

This part of the definition is fortunately easy.
As indicated in \cref{witt-vectors},
a Witt vector over $R$ is an infinite stream of coefficients in $R$,
\[
 (\dots, x_i, \dots, x_2, x_1, x_0), \quad x_i \in R.
\]
This leads to perhaps the simplest definition in our formalization:
\begin{lstlisting}
def witt_vector (p : ℕ) (R : Type*) := ℕ → R
\end{lstlisting}
The argument \lean{p} is not used in the definition,
but \lean{witt_vector} \lean{p} {R} will have a different ring structure
for each \lean{p}.

\subsection{Universal Calculations}
\label{universal-calculations-2}

In \cref{universal-calculations},
we sketched a strategy for proving identities
between operators on the ring of Witt vectors.
This strategy was imprecise,
and as Hazewinkel wrote, it only gives a valid proof if it is a
``universal calculation; that is,
makes no use of any properties beyond those that follow from the axioms
for a unital commutative ring only.''

In the remainder of this paper,
we will use the term ``universal calculation''
in the following precise way:
it is a calculation with \emph{polynomial functions}
on the ring of Witt vectors.
Let us now explain what we mean by a polynomial function.

Many of the operations on $\Witt R$ that we will study
have a polynomial structure to them.
Let $f_R \colon \Witt R \to \Witt R$ be a family of functions
where $R$ ranges over all commutative rings.
In practice, this family is defined by parametrizing over $R$,
so we refer to it as $f$.
We say that $f$ is a \emph{polynomial function} if
there is a family of polynomials $\varphi_n \in \ZZ[X_0, X_1, \ldots]$
such that for every commutative ring $R$
and each $n \in \NN$ and $x = (\ldots x_1, x_0) \in \Witt R$,
\[
  f_R(x)_n = \varphi_n(x_0, x_1, \ldots).
\]

We formalize this as a predicate on the family of functions $f_R$.
\begin{lstlisting}
def is_poly
  (f : Π {R : Type} [comm_ring R], 𝕎 R → 𝕎 R) :
  Prop :=
∃ φ : ℕ → mv_polynomial ℕ ℤ,
  ∀ {R : Type} [comm_ring R] (x : 𝕎 R),
    (f x).coeff = λ n, aeval x.coeff (φ n)
\end{lstlisting}
The square brackets around \lean{comm_ring R} denote that
this is a type class argument.
The function \lean{aeval}
evaluates a multivariate polynomial
given values for the variables in an algebra over the coefficient ring.

The power of this predicate comes from its extensionality principle,
a corollary of \cref{witt-structure-int-exists-unique}.

\begin{slemma}
  \label{poly-ext}
  Let $f, g : \Witt R \to \Witt R$ be polynomial functions,
  witnessed respectively by families of polynomials
  $\phi_n, \psi_n \in \ZZ[\dots, X_1, X_0]$.
  If for all $n \in \NN$ we have
  \[
   W_n(\dots, \phi_1, \phi_0) = W_n(\dots, \psi_1, \psi_0),
  \]
  then $\phi_n = \psi_n$ for all $n \in \NN$,
  and hence $f = g$.
\end{slemma}
In other words, two polynomial functions $f$ and $g$ are equal when
we obtain identical values when
evaluating the Witt polynomials on their underlying polynomials.
The condition
\[
  W_n(\dots, \phi_1, \phi_0) = W_n(\dots, \psi_1, \psi_0)
\]
can be written equivalently as
\begin{equation}
  \label{pre-poly-funext}
  \bind_\varphi (W_n) = \bind_\psi (W_n).
\end{equation}
This is an equality of polynomials with coefficients in $\ZZ$,
and as such it holds exactly when
\[
  \bind_\varphi (W_n)(x_0, x_1, \ldots) = \bind_\psi (W_n)(x_0, x_1, \ldots)
\]
for every sequence of integers $x_i$, $i \in \NN$.

Note that $\bind_\varphi (W_n)(x_0, x_1, \ldots)$
is equal to
\[
  W_n(\dots, \varphi_1(x_0, x_1, \dots), \varphi_0(x_0, x_1, \dots)),
\]
and by assumption
\[
  \varphi_i(x_0, x_1, \ldots) = f(x)_i.
\]
Additionally, recall
the $n$th ghost component (\cref{universal-calculations}):
\[
 w_n \colon \Witt R \to R, \quad x \mapsto W_n(x).
\]
Putting all these pieces together, we see that
\cref{pre-poly-funext} is equivalent to
\[
  \forall x \in \Witt \ZZ, n \in \NN, \quad w_n(f(x)) = w_n(g(x)).
\]

Proving this for all $x \in \Witt R$
for a generic ring $R$
is never harder than proving it for $x \in \Witt \ZZ$,
and the former clearly implies the latter.
The Lean statement of this extensionality principle emphasizes
that the condition reduces the calculation to one over generic rings.
\begin{lstlisting}
lemma is_poly.ext
  {f g : Π {R} [comm_ring R], 𝕎 R → 𝕎 R}
  (hf : is_poly p f) (hg : is_poly p g)
  (heq : ∀ {R} [comm_ring R] (x : 𝕎 R) (n : ℕ),
    ghost_component n (f x) =
      ghost_component n (g x)) :
∀ {R} [comm_ring R] (x : 𝕎 R), f x = g x
\end{lstlisting}

What we have just described is, in fact, the precise version
of the second strategy for proving identities between functions on Witt vectors
that we sketch in \cref{universal-calculations}.
The restricted language that our technique targets
is that of unital commutative rings and morphisms between them.
Essentially, the calculations we carry out in this language
may not depend on features of the specific rings in question:
they may not assume that $p$ is invertible,
that the rings have certain characteristic,
or anything of the sort.

We can thus rephrase our strategy from before:
\paragraph{Strategy 2}
\begin{itemize}
  \item Show that both sides of the identity are given by
   $\NN$-indexed families of polynomial operations
   on the coefficients of Witt vectors.
  \item Show that these polynomial operations are equal.
  \item Use \cref{poly-ext} to reduce this to a computation on ghost components.
 \end{itemize}

Because the ghost component computations invoke only a restricted language,
they tend to be pleasant to carry out,
and are typically provable by the simplifier with little or no extra input.
For example, we apply our strategy to check the relation $F \circ V = [p]$.
The proof in Lean,
when we provide the polynomial structure by hand, is approximately:
\begin{lstlisting}
lemma frobenius_verschiebung (x : 𝕎 R) :
  frobenius (verschiebung x) = x * p :=
is_poly.ext
 ((frobenius_is_poly p).comp verschiebung_is_poly)
 (mul_n_is_poly p p)
 (by ghost_simp) _ _
\end{lstlisting}
The tactic \lean{ghost_simp} does little besides invoke the simplifier with a custom set of lemmas,
proving the goal:
\begin{lstlisting}
∀ (n : ℕ),
  ⇑(ghost_component n)
    (⇑frobenius (⇑verschiebung x)) =
  ⇑(ghost_component n) (x * ↑p)
\end{lstlisting}
We write ``approximately'' because, in fact,
the first argument to \lean{is_poly.ext} can be found automatically as well.
We discuss this in \cref{subsection:auxtac}.

The notion of a polynomial function $\Witt R \to \Witt R$ extends,
in an obvious way, to functions $(\Witt R)^n \to \Witt R$ of any arity.
Addition and multiplication of Witt vectors, for instance,
are polynomial by definition.
Defining binary versions of the predicate, extensionality lemma, and composition rules
further increase the opportunities to use this strategy.

Strategy 2 is powerful and straightforward.
The downside is that it is hard to turn the principle into
a fully generic and flexible machine.
One limitation appears when we consider $\tau$ (Teichm\"uller),
a function $R \to \Witt R$,
which doesn't fit in the framework of $n$-ary functions
from $\Witt R$ to itself.
The lack of a convenient library in \mathlib for the composition of $n$-ary functions
prevents us from using an \lean{is_poly} predicate
on functions of arbitrary arity.
For our current applications, this is no great barrier.
We are able to use Strategy 1 to work around these restrictions when needed.
In the future,
it would be interesting to extend our technique
to make this strategy more widely applicable.

\section{Ring Structure and Other Operations}
\label{section:ring-operations}

Our task now is to define the ring structure on $\Witt R$
and the Teichm\"uller, Verschiebung, Frobenius, and multiplication-by-$n$ operations.
The Teichm\"uller operator will not make an appearance in \cref{section:compare},
but we include it in the interest of
establishing a general interface for Witt vectors in \mathlib.

Our proofs proceed, as much as possible, as universal calculations.
Following Strategy 2 as explained in the previous section
allows many proofs to use essentially the same arguments,
for instance, when we establish the homomorphism properties of the operators.
These arguments are similar enough
that we were able to factor them into short metaprograms,
only slightly more complicated than tactic macros,
that can replicate them with minimal user input.

\subsection{The Ring of Witt Vectors}
\label{subsection:witt-lean}

In \cref{ring-structure} we defined,
for Witt vectors $x = (\dots, x_1, x_0)$ and $y = (\dots, y_1, y_0)$ in $\Witt R$,
\[
\begin{aligned}
 x + y &= (\dots, S_i(x,y), \dots, S_1(x,y), S_0(x,y)) \\
 x \cdot y &= (\dots, P_i(x,y), \dots, P_1(x,y), P_0(x,y)) \\
\end{aligned}
\]
for then-unspecified families of polynomials $S_i$ and~$P_i$.
We obtain these families, which we call \emph{structure polynomials},
by applying \cref{witt-structure-int-exists-unique}
to the bivariate polynomials $X + Y$ and $X \cdot Y$.
The structure polynomial for negation
is obtained similarly with the univariate polynomial $-X$.

In Lean, the unique family of polynomials from \cref{witt-structure-int-exists-unique}
goes by the name \lean{witt_structure_int}.
We define:
\begin{lstlisting}
def witt_add :
  ℕ → mv_polynomial (fin 2 × ℕ) ℤ :=
witt_structure_int p (X 0 + X 1)

def witt_mul :
  ℕ → mv_polynomial (fin 2 × ℕ) ℤ :=
witt_structure_int p (X 0 * X 1)

def witt_neg :
  ℕ → mv_polynomial (fin 1 × ℕ) ℤ :=
witt_structure_int p (-X 0)
\end{lstlisting}

The addition on $\Witt R$ is then defined by letting the $n$th coefficient
be the evaluation of \lean{witt_add n} on the coefficients of $x, y \in \Witt R$.
Multiplication, negation, and the elements $0$ and~$1$ are defined similarly.
\begin{lstlisting}
def eval {k : ℕ}
  (φ : ℕ → mv_polynomial (fin k × ℕ) ℤ)
  (x : fin k → 𝕎 R) : 𝕎 R :=
mk p (λ n, peval (φ n) $ λ i, (x i).coeff)
\end{lstlisting}

\begin{lstlisting}
instance : has_add (𝕎 R) :=
⟨λ x y, eval (witt_add p) ![x, y]⟩
\end{lstlisting}
The function \lean{peval} is simply an uncurried application of \lean{aeval}.
The notation \lean{![x, y]} stands for the function of type \lean{fin 2 → 𝕎 R}
mapping \lean{0} to \lean{x} and \lean{1} to \lean{y}.

To show that these definitions make $\Witt R$ into a commutative ring,
we must check that they satisfy a number of axioms.
Doing this explicitly would be tedious.
We therefore follow Strategy~1, as explained in \cref{universal-calculations}.

Suppose that $f \colon R \to S$ is a function,
both $R$ and $S$ are endowed with $0$, $1$, $+$, $\cdot$, and $-$,
and the function $f$ preserves this structure.
If $f$ is injective, and $S$ satisfies the axioms of a commutative ring,
then so does~$R$.
In Lean this fact is recorded in \lean{function.injective.comm_ring}.
Dually, if $f$ is surjective, and $R$ satisfies the axioms of a commutative ring,
then so does~$S$.
This is \lean{function.surjective.comm_ring}.

For every ring homomorphism $f \colon R \to S$, the map
\[
  \Witt f \colon \Witt R \to \Witt S, \quad (\ldots, x_1, x_0) \mapsto (\ldots, f(x_1), f(x_0))
\]
preserves the ring operations.
We prove this in five lemmas, one lemma for each of $0$, $1$, $+$, $\cdot$, and $-$.
These goals are uniform enough that all can be proved by the same five line tactic script,
which we factor into a tactic macro.
This is not the only place we encounter repetitive goals like this,
and we elaborate on the use of auxiliary tactics in \cref{subsection:auxtac}.

We can then argue as follows that $\Witt R$ is a commutative ring:
\begin{itemize}
 \item
		Recall from \cref{universal-calculations}
  the ring homomorphism
  \[
   w \colon \Witt R \to R^\NN, \quad x \mapsto (W_0(x), W_1(x), \dots),
  \]
  called the ghost map.
  If $p$ is invertible in~$R$,
  then the ghost map is injective.
  So in this case $\Witt R$ is a commutative ring.
 \item If $R = \ZZ[(X_i)_{i \in I}]$ is some polynomial algebra over the integers,
  then we use the natural injection
  \[
   \Witt(\ZZ[(X_i)_{i \in I}]) \to \Witt(\QQ[(X_i)_{i \in I}])
  \]
  and the fact that it preserves the ring operations, as discussed above.
  Since $p$ is invertible in $\QQ[(X_i)_{i \in I}]$,
  we deduce that $\Witt(\ZZ[(X_i)_{i \in I}])$ is a commutative ring
  from \lean{function.injective.comm_ring} and the previous point.
 \item Finally, for arbitrary commutative rings~$R$,
  consider the map $\Witt f$, where $f$ is the natural surjection
  \[
   f \colon \ZZ[(X_r)_{r \in R}] \to R.
  \]
  Since $f$ is surjective, so is $\Witt f$,
  and we can therefore conclude that $\Witt R$ is a commutative ring
  from the fact that $\Witt(\ZZ[(X_r)_{r \in R}])$ is a commutative ring.
\end{itemize}

We have used Strategy~1,
as opposed to Strategy~2 which was explained in \cref{universal-calculations-2},
for two reasons.
\begin{enumerate}[label=(\textit{\roman*})]
 \item With our current approach we deduce all the axioms at once,
  whereas with Strategy~2 we would have to check them one by one.
 \item The associativity axioms refer to ternary operations,
  and we have only formalized the machinery of Strategy~2
  in the unary and binary setting.
  So far, we haven't found a direct use for higher arity versions
  besides the associativity axioms,
  and without a convenient way to uniformly handle $n$-ary versions,
  it was not worth the effort to develop ternary machinery for this single application.
\end{enumerate}

\subsection{Operators}
\label{subsection:operators}

In \cref{ring-structure} we introduced the four operators
Verschiebung ($V$), Frobenius ($F$), scalar multiplication ($[n]$), and Teichm\"uller ($\tau$).
For $x, y \in \Witt R$, these operations satisfy
\begin{align*}
 F \circ V &= [p] \\
 V(x \cdot F(y)) &= V(x) \cdot y
\intertext{and if $R$ has characteristic~$p$}
 F(x) &= (\dots, x_2^p, x_1^p, x_0^p) \\
 [p](x) &= (\dots, x_2^p, x_1^p, x_0^p, 0) \\
 V \circ F &= [p] \\
 p &= (\dots, 0, 0, 1, 0).
\end{align*}
We will need most of these operations and properties in \cref{subsection:isomorphism},
although the Teichm\"uller lift is included only for the sake of completeness.
Teichm\"uller also distinguishes itself as the one operator
whose properties we cannot establish via Strategy~2.
We will show that each of the others is a polynomial function.

\paragraph{Verschiebung}
The definition of the Verschiebung map,
\[
  V \colon \Witt R \to \Witt R, \quad (\dots, x_2, x_1, x_0) \mapsto (\dots, x_2, x_1, x_0, 0),
\]
translates easily to Lean:
\begin{lstlisting}
def verschiebung_fun (x : 𝕎 R) : 𝕎 R :=
mk p (λ n, if n = 0 then 0 else x.coeff (n - 1))
\end{lstlisting}
Its underlying polynomial structure is similarly straightforward:
\begin{lstlisting}
def versch_poly (n : ℕ) : mv_polynomial ℕ ℤ :=
if n = 0 then 0 else X (n-1)
\end{lstlisting}
One lemma, an identity for $\bind_V(W_n)$, is somewhat tedious.
Otherwise, it is routine to show that $V$
is indeed a polynomial function,
respects addition,
is a natural transformation,
interacts with the ghost components.

\paragraph{Multiplication by $n$}
For any $n \in \NN$,
multiplication by $n$ in the ring of Witt vectors
\[
  [n] \colon \Witt R \to \Witt R, \quad x \mapsto n \cdot x
 \]
is a polynomial function,
because it is repeatedly applied addition,
which is polynomial.
The operation needs no definition in Lean
since the coercion \lstinline{ℕ → 𝕎 R} and multiplication on \lstinline{𝕎 R} are known:
it is simply \lstinline{λ x, x * n}.

\paragraph{Frobenius}

The next operator puts up more of a fight.
If $R$ is a ring of characteristic~$p$,
then $f \colon R \to R, r \mapsto r^p$ is a ring endomorphism.
We use this to obtain an endomorphism $\Witt f \colon \Witt R \to \Witt R$,
taking the image of each input coefficient under $f$ (\cref{subsection:witt-lean}).

We claim that $\Witt f$ is a polynomial function,
which unlocks the toolkit of universal calculations,
as described in \cref{universal-calculations-2}.
In addition, we can use those polynomials to define an endomorphism
$F \colon \Witt R \to \Witt R$ for arbitrary rings~$R$
that agrees with $\Witt f$ in the case that $R$ has characteristic~$p$.
Unfortunately we cannot use the machinery of \cref{witt-structure-int-exists-unique}
(\lean{witt_structure_int} in Lean)
to derive these polynomials.
It holds that
\[
  \bind_F (W_n) = W_{n+1},
\]
but to apply \cref{witt-structure-int-exists-unique},
we need this to be a polynomial expression in $W_0, \ldots, W_n$.
Since $W_{n+1}$ contains the variable $X_{n+1}$
it cannot be expressed in terms of the earlier Witt polynomials.

This is a very painful off-by-one error.
Without being able to use the \lean{witt_structure_int} machinery,
we are forced to define the underlying polynomial structure by hand.
The proof that it witnesses that $F$ is a polynomial function
mimics the argument lifting
\cref{witt-structure-rat-exists-unique} (over $\QQ$)
to \cref{witt-structure-int-exists-unique} (over $\ZZ$).
While the high level approach is similar,
the details are different enough that
it is not clear how to unify the calculations.

After establishing that $F$ is polynomial, though,
we are back in the realm of universal calculations.
Further properties of $F$ follow without excess trouble:
for instance, if $x = (\ldots x_1, x_0) \in \Witt R$
and $R$ has characteristic $p$, then
\[
  \left(F\left(x\right)\right)_n = x_n^p
\]
so that $F$ agrees with $\Witt f$ as promised.

\paragraph{Teichm\"uller}

The signature of the Teichm\"uller lift $\tau$
is not the same as the previous operators,
which means we can neither construct it as a polynomial function
nor reason with universal calculations.
Fortunately, its definition
\[
 \tau \colon R \to \Witt R, \quad r \mapsto (\dots, 0, 0, r)
\]
is easy to translate directly.
\begin{lstlisting}
def teichmuller_fun (r : R) : 𝕎 R
| 0 := r
| (n+1) := 0
\end{lstlisting}
After establishing that the $n$th ghost component of $\tau(r)$ is $r^{p^n}$,
it is straightforward to show that $\tau$ is multiplicative and zero-preserving.

While $\tau$ is not needed in \cref{section:compare},
it is an essential part of the Witt vector interface,
and so we define it for the sake of completeness.
It is a multiplicative map inverse to the ring homomorphism
\[
  w_0 \colon \Witt R \to R, \quad (\dots, x_1, x_0) \mapsto x_0.
\]
In \cref{subsection:truncated} we will see
a universal property of $\Witt$,
namely, that it is the projective limit of the rings of truncated Witt vectors.
This shows how to build a ring homomorphism \emph{into} $\Witt R$.
Another universal property of $\Witt$
shows that to build a ring homomorphism \emph{out of} $\Witt k$,
when $k$ is a perfect ring,
it suffices to give suitable values at the Teichm\"uller representatives.
We have not formalized this universal property.




\subsection{Auxiliary Tactics}
\label{subsection:auxtac}

An appealing feature of Lean as a proof assistant
is the easy accessibility of its metaprogramming framework~\cite{EURAM17}.
Lean metaprograms are written in an extension of the language of Lean itself.
With very little syntactic overhead,
these metaprograms can implement tactics
ranging from straightforward macros
to procedures that interact with the parser and environment
in complex ways.

Our adherence to universal calculations leads to a number of proofs
that are identical modulo a few key lemmas or parameters.
A typical example of this is in the previous section,
when we show that the ghost maps respect the ring operations.
The creative step of these proofs is to provide an input polynomial
and the correct arguments
for use by the Witt structure polynomials;
otherwise, the proofs proceed by predictable rewriting.

These predictable proofs fall just outside the scope of a tactic macro,
but can be handled easily by a metaprogram
that parses and inserts arguments.
In the case of the ghost map morphism properties,
a metaprogram that proves all four cases
is only a few lines longer than a direct proof of one case.
We use this approach a number of times while constructing the ring structure on $\Witt R$.
These metaprograms are only used locally,
but let us avoid code duplication and highlight the universality of the proof approach.

We are in an even better position now that the ring structure on $\Witt R$ has been established.
A custom set of simplifier lemmas,
containing the proper universal ghost component equations,
ring homomorphism rules,
and some other glue,
is able to handle the ``predictable rewriting'' step
across a variety of different applications,
in particular when we establish identities between $V$, $F$, and $[n]$.
The proofs of an identity $L = R$ typically takes two steps:
we first use \cref{poly-ext} to reduce the proof to a calculation with ghost components,
which then follows by predictable rewriting.

The first part appears to need some user input,
since \cref{poly-ext} asks for proofs that $L$ and $R$ are polynomial.
But, in fact, these proofs follow predictable patterns as well.
$L$ and $R$ are almost always compositions of atomic functions
that we show to be polynomial at the point of definition.
Establishing that $L$ and $R$ are polynomial amounts to
unpacking the functions' structures
and assembling the appropriate compositionality lemmas in the right order.

This has distinct echoes of the search performed by type class inference.
The \lean{is_poly} predicate does indeed behave much like a type class.
However, the somewhat complicated forms of composition
(unary with binary, binary with two unary, diagonalization)
lead to overly difficult higher order unification problems in type class inference.
We solve this by avoiding generic compositional type class instances
and instead using a metaprogram
to generate specific compositional instances for each atomic function.
When we establish \lean{is_poly frobenius}, for example,
we automatically create two instances:
\begin{lstlisting}
Π (f : 𝕎 R → 𝕎 R) [is_poly f],
  is_poly (λ x, frobenius (f x))

Π (f : 𝕎 R → 𝕎 R → 𝕎 R) [is_poly₂ f],
  is_poly₂ (λ x y, frobenius (f x y))
\end{lstlisting}
The creation of these instances is triggered
by applying an attribute to the proof of \lean{is_poly frobenius}:
\begin{lstlisting}
@[is_poly]
lemma frobenius_is_poly : is_poly frobenius := ...
\end{lstlisting}
The predicate \lstinline{is_poly₂} is a binary version of \lstinline{is_poly}.
Similar instances are made for binary polynomial functions.

We define a small tactic \lean{ghost_calc} that,
when facing a goal $L = R$,
applies the appropriate extensionality lemma
and triggers type class inference to infer the polynomial structure of each side.
With this technique, the user never sees
that \lean{is_poly} is a type class.
The relevant instances are generated under the hood
and applied by \lean{ghost_calc}.

Returning to the example in \cref{universal-calculations-2},
we can now inspect the real proof of the identity $F \circ V = [p]$:
\begin{lstlisting}
lemma frobenius_verschiebung (x : 𝕎 R) :
  frobenius (verschiebung x) = x * p :=
by { ghost_calc x, ghost_simp }
\end{lstlisting}

The call to \lean{ghost_calc} infers the polynomial structure of each side
and applies the (univariate) extensionality lemma.
Applying this lemma changes the goal from an identity in \lstinline{𝕎 R}
to one in Witt vectors over a universally quantified ring;
the ring \lean{R} and vector \lean{x} no longer appear in the goal.
The \lean{ghost_calc} tactic clears these obsolete variables
and introduces new ones with the same names.
In the goal after \lean{ghost_calc},
we perceive an illusion that the ring has not changed:
\begin{lstlisting}
p : ℕ
hp : fact (nat.prime p)
R : Type u_1
R._inst : comm_ring R
x : witt_vector p R
⊢ ∀ (n : ℕ),
    ⇑(ghost_component n)
      (⇑frobenius (⇑verschiebung x)) =
    ⇑(ghost_component n) (x * ↑p)
\end{lstlisting}
As before, the \lean{ghost_simp} set of simplification lemmas
is able to close the remaining goal.
This same proof pattern establishes most of the identities from the previous section,
with minimal or no user input between \lean{ghost_calc} and \lean{ghost_simp}.

\section{Isomorphism with $\Zp$}
\label{section:compare}

So far we have worked with Witt vectors over an arbitrary ring $R$.
As discussed in \cref{witt-fp},
when we specialize $R$ to $\GF{p}$,
we get a ring isomorphic to $\Zp$.
We show this by proving that $\Witt (\GF{p})$ satisfies the same universal property
that we established for $\Zp$ in \cref{subsection:padics:universal}.

\subsection{Truncated Witt Vectors}
\label{subsection:truncated}

Just as we approximated elements of $\Zp$ by truncating all but the rightmost $n$ digits,
so we will truncate Witt vectors to their first $n$ elements.
We define the truncated Witt vectors $\Witt_n R$
to be an $n$-element vector of elements of $R$:
\begin{lstlisting}
def truncated_witt_vector
  (p : ℕ) (n : ℕ) (R : Type*) : Type :=
fin n → R
\end{lstlisting}
The parameter \lean{p : ℕ} is unused in this definition
but will determine the ring structure on this type.
There is a clear map $\Witt R \to \Witt_n R$,
and a map in the other direction appends a stream of 0s
to the end of a truncated vector.
Using these maps, we can lift the ring operations on $\Witt R$ to $\Witt_n R$.
An auxiliary tactic as described in \cref{subsection:auxtac}
establishes that
the truncating map respects the ring operations,
and since this map is surjective,
we can conclude that $\Witt_n R$ is a ring
and the truncating map is a ring homomorphism.

There is another obvious truncation map $\Witt_n R \to \Witt_m R$ for $m \leq n$.
It should come as no surprise that this map is, again, a ring homomorphism.
It additionally composes well with itself and with the full truncating map.
The imaginative reader may now see the lower triangle in \cref{figure:limit},
with $\Witt R$ in the middle and $\Witt_{n+1} R$ and $\Witt_n R$ on the sides.
Indeed, the rest of the diagram follows with no trouble:
given a family of compatible maps $S \to \Witt_n R$,
we can produce a unique map $S \to \Witt R$.

We have said little about the formalization of this section
because there is almost nothing to say.
No argument in this file takes more than a few lines of code:
ring-theoretic machinery and a few simplification lemmas
give us everything practically for free.
It is surprising, then, that
this section contains one of the rare occasions
in which we avoid a ring-theoretic definition.
The type \lean{truncated_witt_vector p n R} could have been represented
as the quotient of \lstinline{𝕎 R}
by the ideal \lstinline{⟨x : 𝕎 R | ∀ i < n, x.coeff i = 0⟩}.
While elegant in principle,
this approach made the definition of coefficients of a truncated Witt vector
rather annoying,
whereas the more direct definition was entirely free of hassle.

\subsection{Constructing the Isomorphism}
\label{subsection:isomorphism}

We now know that $\Zp$ is the projective limit of $\zmod{p^n}$
and $\Witt R$ is the projective limit of $\Witt_n R$.
It is finally time to specialize the arbitrary ring $R$ to $\GF{p}$.
To establish that $\Witt (\GF{p}) \simeq \Zp$,
it suffices by the uniqueness of the projective limit
to show that $\Witt_n (\GF{p}) \simeq \zmod{p^n}$.

It follows immediately that $|\Witt_n (\GF{p})| = p^n$.
A general result shows that a ring $R$ with cardinality $n$ and characteristic $n$
must have a unique isomorphism to $\zmod{n}$,
since both unit elements generate the ring as additive group.
Showing that $\Witt_n (\GF{p})$ has characteristic $n$, though,
takes some machinery:
the proof invokes both the Frobenius and Verschiebung operators
and the identity $F \circ V = [p]$.
Interestingly, this is the first and only time in this development
that we invoke $F$ and $V$,
but developing the theories of these operators
seems to be the shortest path to this result.

\begin{figure}
  \begin{tikzcd}[column sep=huge, row sep=huge]
    \Witt_n (\GF{p}) \arrow["\simeq",leftrightarrow]{r} \arrow{d}[swap]{\text{trunc}} & \zmod{p^n} \arrow{d}{\text{mod}} \\
    \Witt_m (\GF{p}) \arrow["\simeq"',leftrightarrow]{r} & \zmod{p^m}
    \end{tikzcd}
  \caption{The isomorphism $\Witt_n (\GF{p}) \simeq \zmod{p^n}$ commutes with $\text{trunc}$ and $\text{mod}$.}
  \label{figure:commute}
\end{figure}

This isomorphism commutes with the truncation and $\mod$ operators (\cref{figure:commute}).
We then define a family of ring homomorphisms $\Witt (\GF{p}) \to \zmod{p^n}$
by composing this isomorphism with the truncation map from the previous section.
This family is compatible, and thus the universal property of $\Zp$
lifts it to a homomorphism $\Witt (\GF{p}) \to \Zp$.
Similarly, composing the isomorphism with the homomorphism $\Zp \to \zmod{p^n}$
gives a compatible family of homomorphisms $\Zp \to \Witt_n (\GF{p})$,
which the universal property of $\Witt$ lifts to a homomorphism $\Zp \to \Witt (\GF{p})$.
The uniqueness of the limit, and some straightforward rewriting,
let us quickly establish that these maps are inverses,
and thus the two rings are isomorphic.

\begin{lstlisting}
def equiv : 𝕎 (zmod p) ≃+* ℤ_[p] :=
{ to_fun    := to_padic_int p,
  inv_fun   := from_padic_int p,
  left_inv  := from_padic_comp_to_padic_ext _,
  right_inv := to_padic_comp_from_padic_ext _,
  map_mul   := ring_hom.map_mul _,
  map_add   := ring_hom.map_add _ }
\end{lstlisting}

\section{Concluding Thoughts}
\label{section:conclusion}

The \texttt{witt\_vector} directory of our \mathlib branch
contains around 3500 lines of code,
including comments and whitespace,
discounting preliminaries that will be moved to other locations.
Another 1000 lines have been added to the \texttt{padics} directory.
This counts only material corresponding to sections~\ref{section:padics}--\ref{section:compare}.
Many thousands more lines of preliminaries,
especially about multivariate polynomials,
have been or will be incorporated into \mathlib.
While these comparisons are difficult to make scientifically,
we estimate that the 3500 lines correspond to seven dense pages of Hazewinkel~\cite{Haze09}.

To the best of our knowledge,
the ring of Witt vectors has never before been defined in a proof assistant.
Lewis~\cite{Lewi19} surveys the formal developments of $p$-adic numbers
appearing in the literature.
While Pelayo, Voevodsky, and Warren~\cite{Pela15}
take an algebraic approach to defining $\Zp$
that may be amenable to establishing its advanced algebraic properties,
their development does not go beyond the basic ring structure.
Other proof assistant libraries defining $\Zp$
appear to be similarly limited.

Of course, many libraries contain substantial algebraic developments.
In particular,
Coq's Mathematical Components library~\cite{Mahb17}
contains enough group theory to support
Gonthier et al's formalization of the odd order theorem~\cite{Gont13}.
Others, including Cano et al~\cite{Cano16}, 
have enriched the library's ring theory content,
but have focused on computational aspects.
Isabelle's HOL-Algebra library covers many ring-theoretic topics
and an entry by Bordg~\cite{Bord18} in the Archive of Formal Proofs
constructs ring localizations;
the Mizar Mathematical Library also contains a number of articles on ring theory,
including by Korni{\l}owicz and Schwarzweller~\cite{Korn14}
and Watase~\cite{Wata20}.
Avelar et al~\cite{Avel18} describe a formalization of elementary ring theory in PVS.
We are not aware of a development of DVRs or related topics in any of these systems.

Much has been written
about different methods for defining and maintaining
hierarchies of algebraic structures in proof assistants~\cite{Grab16,Mahb13a,Saka20,Spit11}.
In some sense, our project is orthogonal to this literature:
we work at a single fixed point within \mathlib's type class hierarchy.
Nonetheless, there may be some insight here. 
An early attempt at defining Witt vectors in Lean
succumbed to type class searches that were inexplicably long and slow.
A combination of library refactoring
and improved caching in Lean 3's type class inference
have largely resolved these performance issues.
Library refactoring, of course, is rarely fun,
and it is preferable to design hierarchies right the first time.
Tools like Cohen, Sakaguchi, and Tassi's Hierarchy Builder~\cite{Cohe20}
show enormous promise here.
The tabled type class resolution procedure
implemented in Lean 4 by Selsam, Ullrich, and de Moura~\cite{Sels20}
will also allow more flexibility in hierarchy design.

While we were not expecting it from the start,
a very limited amount of Lean metaprogramming
ended up tidying our proof scripts significantly (\cref{subsection:auxtac}).
These tactics did not just shorten the scripts,
but reduced many of them to the point where
the human input---expressions and references to lemmas---was
essentially the same as it would be informally.
We stress that writing these simple tactics requires
no knowledge of the proof assistant's architecture or foundations
and minimal familiarity with the metaprogramming framework.
Mathematical users,
especially those who recognize these repetitive proofs in their own developments,
would spend their time well gaining this minimal familiarity.

Future work on this topic could go in various directions.
Some directions lift extra structure on the base ring $R$
to extra structure on $\Witt R$.
If $R$ is an integral domain of characteristic~$p$,
then $\Witt R$ is an integral domain;
if $k$ is a perfect field of characteristic~$p$,
then $\Witt k$ is a discrete valuation ring.
All the ingredients in the definition of
$p$-adic period ring~$B_{\text{dR}}$ by Fontaine~\cite{Fon94}
are now available in Lean.
Orthogonally, we could define ``big'' Witt vectors,
of which the $p$-typical Witt vectors described here are a quotient.

\begin{acks}
 We thank Jeremy Avigad, Jasmin Blanchette, Kevin Buzzard,
 Sander Dahmen, Gabriel Ebner, and the anonymous reviewers
 for their insightful comments on drafts of this paper.
 We thank the \mathlib community,
 in particular Anne Baanen and Kevin Buzzard,
 for carefully reviewing our formalization
 as it was added to the library.

 The first author receives support from the
 Deutsche For\-schungs Gemeinschaft~(DFG)
 under Graduiertenkolleg~1821
 \emph{(Cohomological Methods in Geometry)}.

 The second author receives support
 from the European Research Council (ERC) under
 the European Union's Horizon 2020 research and innovation
 program (grant agreement No. 713999, Matryoshka)
 and from the Dutch Research Council (NWO) under the
 Vidi program (project No. 016.Vidi.189.037, Lean Forward).
\end{acks}

\bibliographystyle{ACM-Reference-Format}
\bibliography{mathlib-paper}


\begin{thebibliography}{32}


\ifx \showCODEN    \undefined \def \showCODEN     #1{\unskip}     \fi
\ifx \showDOI      \undefined \def \showDOI       #1{#1}\fi
\ifx \showISBNx    \undefined \def \showISBNx     #1{\unskip}     \fi
\ifx \showISBNxiii \undefined \def \showISBNxiii  #1{\unskip}     \fi
\ifx \showISSN     \undefined \def \showISSN      #1{\unskip}     \fi
\ifx \showLCCN     \undefined \def \showLCCN      #1{\unskip}     \fi
\ifx \shownote     \undefined \def \shownote      #1{#1}          \fi
\ifx \showarticletitle \undefined \def \showarticletitle #1{#1}   \fi
\ifx \showURL      \undefined \def \showURL       {\relax}        \fi
\providecommand\bibfield[2]{#2}
\providecommand\bibinfo[2]{#2}
\providecommand\natexlab[1]{#1}
\providecommand\showeprint[2][]{arXiv:#2}

\bibitem[\protect\citeauthoryear{Bordg}{Bordg}{2018}]%
        {Bord18}
\bibfield{author}{\bibinfo{person}{Anthony Bordg}.}
  \bibinfo{year}{2018}\natexlab{}.
\newblock \showarticletitle{The Localization of a Commutative Ring}.
\newblock \bibinfo{journal}{\emph{Archive of Formal Proofs}}
  (\bibinfo{date}{June} \bibinfo{year}{2018}).
\newblock
\showISSN{2150-914x}
\newblock
\shownote{\url{http://isa-afp.org/entries/Localization_Ring.html}, Formal proof
  development.}


\bibitem[\protect\citeauthoryear{Browning}{Browning}{2018}]%
        {browning2018}
\bibfield{author}{\bibinfo{person}{T.~D. Browning}.}
  \bibinfo{year}{2018}\natexlab{}.
\newblock \showarticletitle{How often does the {H}asse principle hold?}
\newblock In \bibinfo{booktitle}{\emph{Algebraic geometry: {S}alt {L}ake {C}ity
  2015}}. \bibinfo{series}{Proc. Sympos. Pure Math.},
  Vol.~\bibinfo{volume}{97}. \bibinfo{publisher}{Amer. Math. Soc., Providence,
  RI}, \bibinfo{pages}{89--102}.
\newblock
\urldef\tempurl%
\url{https://doi.org/10.1090/PSPUM/097.2/01700}
\showDOI{\tempurl}


\bibitem[\protect\citeauthoryear{Buzzard, Commelin, and Massot}{Buzzard
  et~al\mbox{.}}{2020}]%
        {BCM20}
\bibfield{author}{\bibinfo{person}{Kevin Buzzard}, \bibinfo{person}{Johan
  Commelin}, {and} \bibinfo{person}{Patrick Massot}.}
  \bibinfo{year}{2020}\natexlab{}.
\newblock \showarticletitle{Formalising Perfectoid Spaces}. In
  \bibinfo{booktitle}{\emph{Proceedings of the 9th ACM SIGPLAN International
  Conference on Certified Programs and Proofs}} (New Orleans, LA, USA)
  \emph{(\bibinfo{series}{CPP 2020})}. \bibinfo{publisher}{Association for
  Computing Machinery}, \bibinfo{address}{New York, NY, USA},
  \bibinfo{pages}{299–312}.
\newblock
\showISBNx{9781450370974}
\urldef\tempurl%
\url{https://doi.org/10.1145/3372885.3373830}
\showDOI{\tempurl}


\bibitem[\protect\citeauthoryear{Cano, Cohen, D\'en\`es, M\"ortberg, and
  Siles}{Cano et~al\mbox{.}}{2016}]%
        {Cano16}
\bibfield{author}{\bibinfo{person}{Guillaume Cano}, \bibinfo{person}{Cyril
  Cohen}, \bibinfo{person}{Maxime D\'en\`es}, \bibinfo{person}{Anders
  M\"ortberg}, {and} \bibinfo{person}{Vincent Siles}.}
  \bibinfo{year}{2016}\natexlab{}.
\newblock \showarticletitle{Formalized linear algebra over elementary divisor
  rings in Coq}.
\newblock \bibinfo{journal}{\emph{Logical Methods in Computer Science}}
  \bibinfo{volume}{12}, \bibinfo{number}{2} (\bibinfo{date}{Jun}
  \bibinfo{year}{2016}).
\newblock
\showISSN{1860-5974}
\urldef\tempurl%
\url{https://doi.org/10.2168/lmcs-12(2:7)2016}
\showDOI{\tempurl}


\bibitem[\protect\citeauthoryear{Cohen, Sakaguchi, and Tassi}{Cohen
  et~al\mbox{.}}{2020}]%
        {Cohe20}
\bibfield{author}{\bibinfo{person}{Cyril Cohen}, \bibinfo{person}{Kazuhiko
  Sakaguchi}, {and} \bibinfo{person}{Enrico Tassi}.}
  \bibinfo{year}{2020}\natexlab{}.
\newblock \bibinfo{title}{{Hierarchy Builder: algebraic hierarchies made easy
  in Coq with Elpi}}.  (\bibinfo{date}{Feb.} \bibinfo{year}{2020}).
\newblock
\urldef\tempurl%
\url{https://doi.org/10.4230/LIPIcs.CVIT.2016.23}
\showDOI{\tempurl}


\bibitem[\protect\citeauthoryear{da~Silva, de~Lima, and Galdino}{da~Silva
  et~al\mbox{.}}{2018}]%
        {Avel18}
\bibfield{author}{\bibinfo{person}{Andr{\'{e}}ia B.~Avelar da Silva},
  \bibinfo{person}{Thaynara~Arielly de Lima}, {and}
  \bibinfo{person}{Andr{\'{e}}~Luiz Galdino}.} \bibinfo{year}{2018}\natexlab{}.
\newblock \showarticletitle{Formalizing Ring Theory in {PVS}}. In
  \bibinfo{booktitle}{\emph{Interactive Theorem Proving - 9th International
  Conference, {ITP} 2018, Held as Part of the Federated Logic Conference, FloC
  2018, Oxford, UK, July 9-12, 2018, Proceedings}}
  \emph{(\bibinfo{series}{Lecture Notes in Computer Science},
  Vol.~\bibinfo{volume}{10895})}, \bibfield{editor}{\bibinfo{person}{Jeremy
  Avigad} {and} \bibinfo{person}{Assia Mahboubi}} (Eds.).
  \bibinfo{publisher}{Springer}, \bibinfo{pages}{40--47}.
\newblock
\urldef\tempurl%
\url{https://doi.org/10.1007/978-3-319-94821-8\_3}
\showDOI{\tempurl}


\bibitem[\protect\citeauthoryear{Dahmen, H{\"o}lzl, and Lewis}{Dahmen
  et~al\mbox{.}}{2019}]%
        {Dahm19}
\bibfield{author}{\bibinfo{person}{Sander~R. Dahmen}, \bibinfo{person}{Johannes
  H{\"o}lzl}, {and} \bibinfo{person}{Robert~Y. Lewis}.}
  \bibinfo{year}{2019}\natexlab{}.
\newblock \showarticletitle{{Formalizing the Solution to the Cap Set Problem}}.
  In \bibinfo{booktitle}{\emph{10th International Conference on Interactive
  Theorem Proving (ITP 2019)}} \emph{(\bibinfo{series}{Leibniz International
  Proceedings in Informatics (LIPIcs)}, Vol.~\bibinfo{volume}{141})},
  \bibfield{editor}{\bibinfo{person}{John Harrison}, \bibinfo{person}{John
  O'Leary}, {and} \bibinfo{person}{Andrew Tolmach}} (Eds.).
  \bibinfo{publisher}{Schloss Dagstuhl--Leibniz-Zentrum fuer Informatik},
  \bibinfo{address}{Dagstuhl, Germany}, \bibinfo{pages}{15:1--15:19}.
\newblock
\showISBNx{978-3-95977-122-1}
\showISSN{1868-8969}
\urldef\tempurl%
\url{https://doi.org/10.4230/LIPIcs.ITP.2019.15}
\showDOI{\tempurl}


\bibitem[\protect\citeauthoryear{Ebner, Ullrich, Roesch, Avigad, and
  de~Moura}{Ebner et~al\mbox{.}}{2017}]%
        {EURAM17}
\bibfield{author}{\bibinfo{person}{Gabriel Ebner}, \bibinfo{person}{Sebastian
  Ullrich}, \bibinfo{person}{Jared Roesch}, \bibinfo{person}{Jeremy Avigad},
  {and} \bibinfo{person}{Leonardo de Moura}.} \bibinfo{year}{2017}\natexlab{}.
\newblock \showarticletitle{A metaprogramming framework for formal
  verification}.
\newblock \bibinfo{journal}{\emph{{PACMPL}}} \bibinfo{volume}{1},
  \bibinfo{number}{{ICFP}} (\bibinfo{year}{2017}),
  \bibinfo{pages}{34:1--34:29}.
\newblock
\urldef\tempurl%
\url{https://doi.org/10.1145/3110278}
\showDOI{\tempurl}


\bibitem[\protect\citeauthoryear{Fontaine}{Fontaine}{1994}]%
        {Fon94}
\bibfield{author}{\bibinfo{person}{Jean-Marc Fontaine}.}
  \bibinfo{year}{1994}\natexlab{}.
\newblock \showarticletitle{Le corps des p\'{e}riodes {$p$}-adiques}.
\newblock Number 223. \bibinfo{pages}{59--111}.
\newblock
\showISSN{0303-1179}
\newblock
\shownote{With an appendix by Pierre Colmez, P\'{e}riodes $p$-adiques
  (Bures-sur-Yvette, 1988).}


\bibitem[\protect\citeauthoryear{Gonthier, Asperti, Avigad, Bertot, Cohen,
  Garillot, Roux, Mahboubi, O'Connor, Biha, Pasca, Rideau, Solovyev, Tassi, and
  Th{\'{e}}ry}{Gonthier et~al\mbox{.}}{2013}]%
        {Gont13}
\bibfield{author}{\bibinfo{person}{Georges Gonthier}, \bibinfo{person}{Andrea
  Asperti}, \bibinfo{person}{Jeremy Avigad}, \bibinfo{person}{Yves Bertot},
  \bibinfo{person}{Cyril Cohen}, \bibinfo{person}{Fran{\c{c}}ois Garillot},
  \bibinfo{person}{St{\'{e}}phane~Le Roux}, \bibinfo{person}{Assia Mahboubi},
  \bibinfo{person}{Russell O'Connor}, \bibinfo{person}{Sidi~Ould Biha},
  \bibinfo{person}{Ioana Pasca}, \bibinfo{person}{Laurence Rideau},
  \bibinfo{person}{Alexey Solovyev}, \bibinfo{person}{Enrico Tassi}, {and}
  \bibinfo{person}{Laurent Th{\'{e}}ry}.} \bibinfo{year}{2013}\natexlab{}.
\newblock \showarticletitle{A Machine-Checked Proof of the Odd Order Theorem}.
  In \bibinfo{booktitle}{\emph{{ITP} 2013}}. \bibinfo{pages}{163--179}.
\newblock
\urldef\tempurl%
\url{https://doi.org/10.1007/978-3-642-39634-2\_14}
\showDOI{\tempurl}


\bibitem[\protect\citeauthoryear{Gouv\^{e}a}{Gouv\^{e}a}{1997}]%
        {Gouv97}
\bibfield{author}{\bibinfo{person}{Fernando~Q. Gouv\^{e}a}.}
  \bibinfo{year}{1997}\natexlab{}.
\newblock \bibinfo{booktitle}{\emph{{$p$}-adic Numbers}
  (\bibinfo{edition}{second} ed.)}.
\newblock \bibinfo{publisher}{Springer, Berlin}. vi+298 pages.
\newblock
\showISBNx{3-540-62911-4}
\urldef\tempurl%
\url{https://doi.org/10.1007/978-3-642-59058-0}
\showDOI{\tempurl}


\bibitem[\protect\citeauthoryear{Grabowski, Kornilowicz, and
  Schwarzweller}{Grabowski et~al\mbox{.}}{2016}]%
        {Grab16}
\bibfield{author}{\bibinfo{person}{Adam Grabowski}, \bibinfo{person}{Artur
  Kornilowicz}, {and} \bibinfo{person}{Christoph Schwarzweller}.}
  \bibinfo{year}{2016}\natexlab{}.
\newblock \showarticletitle{On algebraic hierarchies in mathematical repository
  of Mizar}. In \bibinfo{booktitle}{\emph{Proceedings of the 2016 Federated
  Conference on Computer Science and Information Systems, FedCSIS 2016,
  Gda{\'{n}}sk, Poland, September 11-14, 2016.}} \bibinfo{pages}{363--371}.
\newblock
\urldef\tempurl%
\url{https://doi.org/10.15439/2016F520}
\showDOI{\tempurl}


\bibitem[\protect\citeauthoryear{Hales, Adams, Bauer, Dang, Harrison, Hoang,
  Kaliszyk, Magron, McLaughlin, Nguyen, Nguyen, Nipkow, Obua, Pleso, Rute,
  Solovyev, Ta, Tran, Trieu, Urban, Vu, and Zumkeller}{Hales
  et~al\mbox{.}}{2017}]%
        {Hales15}
\bibfield{author}{\bibinfo{person}{Thomas~C. Hales}, \bibinfo{person}{Mark
  Adams}, \bibinfo{person}{Gertrud Bauer}, \bibinfo{person}{Dat~Tat Dang},
  \bibinfo{person}{John Harrison}, \bibinfo{person}{Truong~Le Hoang},
  \bibinfo{person}{Cezary Kaliszyk}, \bibinfo{person}{Victor Magron},
  \bibinfo{person}{Sean McLaughlin}, \bibinfo{person}{Thang~Tat Nguyen},
  \bibinfo{person}{Truong~Quang Nguyen}, \bibinfo{person}{Tobias Nipkow},
  \bibinfo{person}{Steven Obua}, \bibinfo{person}{Joseph Pleso},
  \bibinfo{person}{Jason~M. Rute}, \bibinfo{person}{Alexey Solovyev},
  \bibinfo{person}{An~Hoai~Thi Ta}, \bibinfo{person}{Trung~Nam Tran},
  \bibinfo{person}{Diep~Thi Trieu}, \bibinfo{person}{Josef Urban},
  \bibinfo{person}{Ky~Khac Vu}, {and} \bibinfo{person}{Roland Zumkeller}.}
  \bibinfo{year}{2017}\natexlab{}.
\newblock \showarticletitle{A formal proof of the Kepler conjecture}.
\newblock \bibinfo{journal}{\emph{Forum of Mathematics, Pi}}
  \bibinfo{volume}{5} (\bibinfo{year}{2017}), \bibinfo{pages}{e2}.
\newblock
\urldef\tempurl%
\url{https://doi.org/10.1017/fmp.2017.1}
\showDOI{\tempurl}


\bibitem[\protect\citeauthoryear{Hazewinkel}{Hazewinkel}{2009}]%
        {Haze09}
\bibfield{author}{\bibinfo{person}{Michiel Hazewinkel}.}
  \bibinfo{year}{2009}\natexlab{}.
\newblock \showarticletitle{Witt vectors. Part 1}.
\newblock \bibinfo{journal}{\emph{Handbook of Algebra}} (\bibinfo{year}{2009}),
  \bibinfo{pages}{319–472}.
\newblock
\showISBNx{9780444532572}
\showISSN{1570-7954}
\urldef\tempurl%
\url{https://doi.org/10.1016/s1570-7954(08)00207-6}
\showDOI{\tempurl}


\bibitem[\protect\citeauthoryear{Korni{\l}owicz and
  Schwarzweller}{Korni{\l}owicz and Schwarzweller}{2014}]%
        {Korn14}
\bibfield{author}{\bibinfo{person}{Artur Korni{\l}owicz} {and}
  \bibinfo{person}{Christoph Schwarzweller}.} \bibinfo{year}{2014}\natexlab{}.
\newblock \showarticletitle{{T}he First Isomorphism Theorem and Other
  Properties of Rings}.
\newblock \bibinfo{journal}{\emph{Formalized Mathematics}}
  \bibinfo{volume}{22}, \bibinfo{number}{{\bf 4}} (\bibinfo{year}{2014}),
  \bibinfo{pages}{291--301}.
\newblock
\showISSN{1426-2630}
\urldef\tempurl%
\url{https://doi.org/10.2478/forma-2014-0029}
\showDOI{\tempurl}


\bibitem[\protect\citeauthoryear{Lech}{Lech}{1953}]%
        {lech1953}
\bibfield{author}{\bibinfo{person}{Christer Lech}.}
  \bibinfo{year}{1953}\natexlab{}.
\newblock \showarticletitle{A note on recurring series}.
\newblock \bibinfo{journal}{\emph{Ark. Mat.}} \bibinfo{volume}{2},
  \bibinfo{number}{5} (\bibinfo{date}{08} \bibinfo{year}{1953}),
  \bibinfo{pages}{417--421}.
\newblock
\urldef\tempurl%
\url{https://doi.org/10.1007/BF02590997}
\showDOI{\tempurl}


\bibitem[\protect\citeauthoryear{Lewis}{Lewis}{2019}]%
        {Lewi19}
\bibfield{author}{\bibinfo{person}{Robert~Y. Lewis}.}
  \bibinfo{year}{2019}\natexlab{}.
\newblock \showarticletitle{A formal proof of {H}ensel's lemma over the
  $p$-adic integers}. In \bibinfo{booktitle}{\emph{Proceedings of the 8th {ACM}
  {SIGPLAN} International Conference on Certified Programs and Proofs, {CPP}
  2019, Cascais, Portugal, January 14-15, 2019}}. \bibinfo{pages}{15--26}.
\newblock
\urldef\tempurl%
\url{https://doi.org/10.1145/3293880.3294089}
\showDOI{\tempurl}


\bibitem[\protect\citeauthoryear{Lewis and Madelaine}{Lewis and
  Madelaine}{2019}]%
        {Lewi20}
\bibfield{author}{\bibinfo{person}{Robert~Y. Lewis} {and}
  \bibinfo{person}{Paul-Nicolas Madelaine}.} \bibinfo{year}{2019}\natexlab{}.
\newblock \showarticletitle{Simplifying Casts and Coercions}. In
  \bibinfo{booktitle}{\emph{PAAR 2020: Seventh Workshop on Practical Aspects of
  Automated Reasoning, June 29–30, 2020, Paris, France (virtual)}}.
  \bibinfo{pages}{53--62}.
\newblock
\urldef\tempurl%
\url{http://ceur-ws.org/Vol-2752/paper4.pdf}
\showURL{%
\tempurl}


\bibitem[\protect\citeauthoryear{Mahboubi and Tassi}{Mahboubi and
  Tassi}{2013}]%
        {Mahb13a}
\bibfield{author}{\bibinfo{person}{Assia Mahboubi} {and}
  \bibinfo{person}{Enrico Tassi}.} \bibinfo{year}{2013}\natexlab{}.
\newblock \showarticletitle{Canonical Structures for the Working Coq User}. In
  \bibinfo{booktitle}{\emph{Interactive Theorem Proving - 4th International
  Conference, {ITP} 2013, Rennes, France, July 22-26, 2013. Proceedings}}
  \emph{(\bibinfo{series}{Lecture Notes in Computer Science},
  Vol.~\bibinfo{volume}{7998})}, \bibfield{editor}{\bibinfo{person}{Sandrine
  Blazy}, \bibinfo{person}{Christine Paulin{-}Mohring}, {and}
  \bibinfo{person}{David Pichardie}} (Eds.). \bibinfo{publisher}{Springer},
  \bibinfo{pages}{19--34}.
\newblock
\urldef\tempurl%
\url{https://doi.org/10.1007/978-3-642-39634-2\_5}
\showDOI{\tempurl}


\bibitem[\protect\citeauthoryear{Mahboubi and Tassi}{Mahboubi and
  Tassi}{2020}]%
        {Mahb17}
\bibfield{author}{\bibinfo{person}{Assia Mahboubi} {and}
  \bibinfo{person}{Enrico Tassi}.} \bibinfo{year}{2020}\natexlab{}.
\newblock \bibinfo{booktitle}{\emph{Mathematical Components}}.
\newblock \bibinfo{publisher}{Zenodo}.
\newblock
\urldef\tempurl%
\url{https://doi.org/10.5281/zenodo.4282710}
\showDOI{\tempurl}


\bibitem[\protect\citeauthoryear{\mathlibcommunity}{\mathlibcommunity}{2020}]%
        {mathlib20}
\bibfield{author}{\bibinfo{person}{\mathlibcommunity}.}
  \bibinfo{year}{2020}\natexlab{}.
\newblock \showarticletitle{The {L}ean Mathematical Library}. In
  \bibinfo{booktitle}{\emph{CPP}} (New Orleans, LA, USA).
  \bibinfo{publisher}{ACM}, \bibinfo{address}{New York, NY, USA},
  \bibinfo{pages}{367–381}.
\newblock
\showISBNx{9781450370974}
\urldef\tempurl%
\url{https://doi.org/10.1145/3372885.3373824}
\showDOI{\tempurl}


\bibitem[\protect\citeauthoryear{McCallum and Poonen}{McCallum and
  Poonen}{2012}]%
        {mccallum2012}
\bibfield{author}{\bibinfo{person}{William McCallum} {and}
  \bibinfo{person}{Bjorn Poonen}.} \bibinfo{year}{2012}\natexlab{}.
\newblock \showarticletitle{The method of {C}habauty and {C}oleman}.
\newblock In \bibinfo{booktitle}{\emph{Explicit methods in number theory}}.
  \bibinfo{series}{Panor. Synth\`eses}, Vol.~\bibinfo{volume}{36}.
  \bibinfo{publisher}{Soc. Math. France, Paris}, \bibinfo{pages}{99--117}.
\newblock


\bibitem[\protect\citeauthoryear{Pelayo, Voevodsky, and Warren}{Pelayo
  et~al\mbox{.}}{2015}]%
        {Pela15}
\bibfield{author}{\bibinfo{person}{Álvaro Pelayo}, \bibinfo{person}{Vladimir
  Voevodsky}, {and} \bibinfo{person}{Michael~A. Warren}.}
  \bibinfo{year}{2015}\natexlab{}.
\newblock \showarticletitle{A univalent formalization of the p-adic numbers}.
\newblock \bibinfo{journal}{\emph{Mathematical Structures in Computer Science}}
  \bibinfo{volume}{25}, \bibinfo{number}{5} (\bibinfo{year}{2015}),
  \bibinfo{pages}{1147–1171}.
\newblock
\urldef\tempurl%
\url{https://doi.org/10.1017/S0960129514000541}
\showDOI{\tempurl}


\bibitem[\protect\citeauthoryear{Sakaguchi}{Sakaguchi}{2020}]%
        {Saka20}
\bibfield{author}{\bibinfo{person}{Kazuhiko Sakaguchi}.}
  \bibinfo{year}{2020}\natexlab{}.
\newblock \bibinfo{title}{Validating Mathematical Structures}.
\newblock \bibinfo{howpublished}{{arXiv}}.
\newblock
\showeprint[arxiv]{2002.00620}~[cs.PL]
\urldef\tempurl%
\url{https://arxiv.org/abs/2002.00620}
\showURL{%
\tempurl}


\bibitem[\protect\citeauthoryear{Schmid}{Schmid}{1936}]%
        {Schm36}
\bibfield{author}{\bibinfo{person}{Hermann~Ludwig Schmid}.}
  \bibinfo{year}{1936}\natexlab{}.
\newblock \showarticletitle{Zyklische algebraische Funktionenkörper vom Grade
  $p^n$ über endlichem Konstantenkörper der Charakteristik $p$.}
\newblock \bibinfo{journal}{\emph{Journal für die reine und angewandte
  Mathematik}} \bibinfo{volume}{1936}, \bibinfo{number}{175}
  (\bibinfo{year}{1936}), \bibinfo{pages}{108 -- 123}.
\newblock
\urldef\tempurl%
\url{https://doi.org/10.1515/crll.1936.175.108}
\showDOI{\tempurl}


\bibitem[\protect\citeauthoryear{Selsam, Ullrich, and de~Moura}{Selsam
  et~al\mbox{.}}{2020}]%
        {Sels20}
\bibfield{author}{\bibinfo{person}{Daniel Selsam}, \bibinfo{person}{Sebastian
  Ullrich}, {and} \bibinfo{person}{Leonardo de Moura}.}
  \bibinfo{year}{2020}\natexlab{}.
\newblock \bibinfo{title}{Tabled Typeclass Resolution}.
\newblock
\newblock
\showeprint[arxiv]{2001.04301}~[cs.PL]


\bibitem[\protect\citeauthoryear{Spitters and van~der Weegen}{Spitters and
  van~der Weegen}{2011}]%
        {Spit11}
\bibfield{author}{\bibinfo{person}{Bas Spitters} {and} \bibinfo{person}{Eelis
  van~der Weegen}.} \bibinfo{year}{2011}\natexlab{}.
\newblock \showarticletitle{Type classes for mathematics in type theory}.
\newblock \bibinfo{journal}{\emph{Mathematical Structures in Computer Science}}
  \bibinfo{volume}{21}, \bibinfo{number}{4} (\bibinfo{year}{2011}),
  \bibinfo{pages}{795--825}.
\newblock
\urldef\tempurl%
\url{https://doi.org/10.1017/S0960129511000119}
\showDOI{\tempurl}


\bibitem[\protect\citeauthoryear{Strickland and Bellumat}{Strickland and
  Bellumat}{2019}]%
        {Stri19}
\bibfield{author}{\bibinfo{person}{Neil Strickland} {and}
  \bibinfo{person}{Nicola Bellumat}.} \bibinfo{year}{2019}\natexlab{}.
\newblock \bibinfo{title}{Iterated chromatic localisation}.
\newblock
\newblock
\showeprint[arxiv]{1907.07801}~[math.AT]


\bibitem[\protect\citeauthoryear{Wadler and Blott}{Wadler and Blott}{1989}]%
        {Wadl89}
\bibfield{author}{\bibinfo{person}{Philip Wadler} {and}
  \bibinfo{person}{Stephen Blott}.} \bibinfo{year}{1989}\natexlab{}.
\newblock \showarticletitle{How to Make ad-hoc Polymorphism Less ad-hoc}. In
  \bibinfo{booktitle}{\emph{Proceedings of {POPL} 1989}}.
  \bibinfo{pages}{60--76}.
\newblock
\urldef\tempurl%
\url{https://doi.org/10.1145/75277.75283}
\showDOI{\tempurl}


\bibitem[\protect\citeauthoryear{Watase}{Watase}{2020}]%
        {Wata20}
\bibfield{author}{\bibinfo{person}{Yasushige Watase}.}
  \bibinfo{year}{2020}\natexlab{}.
\newblock \showarticletitle{{R}ings of {F}ractions and {L}ocalization}.
\newblock \bibinfo{journal}{\emph{Formalized Mathematics}}
  \bibinfo{volume}{28}, \bibinfo{number}{{\bf 1}} (\bibinfo{year}{2020}),
  \bibinfo{pages}{79--87}.
\newblock
\showISSN{1426-2630}
\urldef\tempurl%
\url{https://doi.org/10.2478/forma-2020-0006}
\showDOI{\tempurl}


\bibitem[\protect\citeauthoryear{Wiedijk}{Wiedijk}{2007}]%
        {Wied07}
\bibfield{author}{\bibinfo{person}{Freek Wiedijk}.}
  \bibinfo{year}{2007}\natexlab{}.
\newblock \bibinfo{title}{The QED Manifesto Revisited}.
\newblock
\newblock


\bibitem[\protect\citeauthoryear{Witt}{Witt}{1937}]%
        {Witt37}
\bibfield{author}{\bibinfo{person}{E. Witt}.} \bibinfo{year}{1937}\natexlab{}.
\newblock \showarticletitle{Zyklische K{\"o}rper und Algebren der
  Charakteristik $p$ vom Grad $p^n$. Struktur diskret bewerteter perfekter
  K{\"o}rper mit vollkommenem Restklassenk{\"o}rper der Charakteristik $p$.}
\newblock \bibinfo{journal}{\emph{Journal f{\"u}r die reine und angewandte
  Mathematik (Crelles Journal)}}  \bibinfo{volume}{1937}
  (\bibinfo{year}{1937}), \bibinfo{pages}{126 -- 140}.
\newblock
\urldef\tempurl%
\url{https://doi.org/10.1515/crll.1937.176.126}
\showDOI{\tempurl}


\end{thebibliography}


\end{document}